\documentclass[lettersize,journal]{IEEEtran}
\usepackage{amsmath,amsfonts}
\usepackage{algorithmic}
\usepackage{algorithm}
\usepackage{array}
\usepackage{textcomp}
\usepackage{stfloats}
\usepackage{comment}
\usepackage{xcolor}
\usepackage{bm}
\usepackage{url}
\usepackage{verbatim}
\usepackage{graphicx}
\usepackage{subcaption}
\usepackage{cite}
\title{A Data-Informed Local Subspaces Method for Error-Bounded Lossy Compression of Large-Scale Scientific Datasets}

\author{\IEEEauthorblockN{Arshan Khan\IEEEauthorrefmark{1},
Rohit Deshmukh\IEEEauthorrefmark{2} \\ 
\IEEEauthorblockA{Department of Mechanical and Aerospace Engineering, \\
University of Central Florida \\
Orlando, FL, USA} \\
\vspace{10 pt}
Ben O'Neill\IEEEauthorrefmark{3} \\
\IEEEauthorblockA{RNET Technologies Inc.\\
Dayton, OH, USA} \\
\vspace{10 pt}
Email: \IEEEauthorrefmark{1}arshan.khan@ucf.edu,\\
\IEEEauthorrefmark{2}rohit.deshmukh@ucf.edu (corresponding author),\\
\IEEEauthorrefmark{3}boneill@rnet-tech.com}}

\begin{document}
\maketitle

\begin{abstract}
The growing volume of scientific simulation data presents a significant challenge for storage and transfer. Error-bounded lossy compression has emerged as a critical solution for mitigating these challenges, providing a means to reduce data size while ensuring that reconstructed data remains valid for scientific analysis. In this paper, we present a data-driven scientific data compressor, called Discontinuous Data-informed Local Subspaces (Discontinuous DLS), to improve compression-to-error ratios over data-agnostic compressors. This error-bounded compressor leverages localized spatial and temporal subspaces, informed by the underlying data structure, to enhance compression efficiency and preserve key features. The presented technique is flexible and applicable to a wide range of scientific data, including fluid dynamics, environmental simulations, and other high-dimensional, time-dependent datasets. We describe the core principles of the method and demonstrate its ability to significantly reduce storage requirements without compromising critical data fidelity. The technique is implemented in a distributed computing environment using MPI, and its performance is evaluated against state-of-the-art error-bounded compression methods in terms of compression ratio and reconstruction accuracy. This study highlights discontinuous DLS as a promising approach for large-scale scientific data compression in high-performance computing environments, providing a robust solution for managing the growing data demands of modern scientific simulations.
\end{abstract}


\section{Introduction}

The rapid advancement of high-performance computing (HPC) has facilitated the generation of increasingly large and complex scientific datasets across various fields, including climate modeling, aerodynamics, genomics, and materials science. With the growing integration of heterogeneous computing architectures, which combine traditional CPUs with accelerators such as GPUs, researchers can simulate and analyze data at an unprecedented scale. However, as computational power continues to increase, the ability to efficiently store, manage, and process these massive datasets becomes a significant challenge. In particular, the demand for memory and storage capacity is outpacing available resources, leading to bottlenecks in data transfer and retrieval \cite{Cappello2019, Zivanovic2017}.
To mitigate these issues, lossy data compression has emerged as a critical tool. Unlike lossless techniques such as Run-Length Encoding (RLE), Huffman Coding, Shannon-Fano Encoding, and Arithmetic Coding, which offer only modest reductions, error-bounded lossy compressors can achieve orders-of-magnitude compression by trading off exact reconstruction for tolerable numerical errors \cite{Delaunay2019,Sharma2017}.
%
%
Over the last decade, several error-bounded lossy compressors have been developed for scientific data. ISABELA \cite{Lakshminarasimhan2013}, a dictionary-based scheme, uses B-spline fitting followed by quantization and entropy encoding. It is effective for smooth and monotonic datasets but experiences performance degradation on irregular data, partly due to increased sorting overhead \cite{Liang2018b, Lakshminarasimhan2013}. SZ \cite{Di2016, Liang2018, Tao2017}, one of the most widely adopted lossy compressors, employs predictive modeling with configurable error bounds and various predictors (Lorenzo, regression, and pattern), offering strong performance on structured grid data with pointwise and absolute target error bounds. ZFP \cite{Lindstrom2024} adopts a transform-based approach with embedded coding and shared-exponent alignment, offering fast compression with high accuracy, although it lacks strict control over maximum error\cite{Cappello2019,Lindstrom2014}.
Meanwhile, MGARD \cite{Ainsworth2018, Ainsworth2019} uses a multilevel decomposition framework based on finite element interpolation and wavelet theory to provide guarantees of norm error. Its enhanced version, MGARD+ \cite{Liang2022}, has demonstrated significantly improved compression ratios across a range of scientific domains, including climate, quantum chemistry, and astrophysics. However, the method's complexity and performance cost are higher compared to SZ or ZFP \cite{Liang2022}. TTHRESH applies Tucker decomposition and bit-plane coding on tensor data, achieving strong compression for large tolerances, but with high computational cost and limited support for pointwise error control \cite{Ballester2020}.

A common feature of these methods is that they are closer to data-agnosticism than data-driven,i.e., their predictor function types, such as wavelets, splines, and polynomials, are predefined irrespective of the data. We hypothesize that a data-driven predictor function --- one derived from the target data itself --- will result in improved compression ratios. To this end, we propose a family of Data-informed Local Subspaces (DLS) based lossy compression algorithms. Origin of these algorithms lies in the Generalized Finite Element Method (GFEM) \cite{Duarte2000, Kim2010,Hara2009}, which allows incorporation of arbitrary numerical and analytical functions within low-order finite element spaces. Deshmukh {\it et al.} \cite{Deshmukh2020} developed the $C^0$-Continous form of the DLS algorithm by learning these numerical functions directly from the raw target data by spatially-locally sampling the global solution fields. In the context of a lossy compressor, the data-derived enrichment functions are viewed as "data-driven predictor functions". Deshmukh {\it et al.} \cite{Deshmukh2020} demonstrated efficient representations of highly turbulent fluid flow fields. This DLS approach, outlined in \cite{Deshmukh2020a}, hereafter referred to as $C^0$-DLS and/or vanilla DLS, uses the GFEM to maintain $C^0$ continuity, ensuring smooth representations of the data. However, it suffers from scalability limitations, particularly for large and complex datasets, as it requires additional computational resources to achieve lower error levels. 

In this work, we developed an error-bounded, discontinuous DLS approach that relaxes $C^0$ continuity constraints while systematically controlling localized error. Relaxing this constraint makes the method naturally scalable, straightforward to parallelize, and flexible with respect to user-prescribed error tolerances. We demonstrate the method on a three-dimensional incompressible turbulent flow past a stationary cylinder at $Re = 10^5$ \cite{Deshmukh2020a}. A comprehensive evaluation is provided, examining the algorithmic formulation, the impact of compression-induced information loss on key physical features of the flow, and the associated computational costs. The remainder of the paper is structured as follows. Section \S 2 briefly reviews the vanilla $C^0$-DLS formulation and describes the proposed discontinuous DLS formulation. Section \S 3 presents a comparison with other state-of-the-art error-bounded lossy compressors.  Section \S 4, \S 5 and \S 6 discuss the Compression results and the preservation of physical fidelity in the considered dataset. Section \S 7 concludes the paper and outlines directions for future work.

\section{Mathematical Background}

\subsection{Vanilla $C^0$-DLS Compression Algorithm}

The compression strategy developed in this paper is motivated by ideas from the generalized finite element method (GFEM). In GFEM, accuracy is improved not by globally refining the mesh, but by locally enriching the approximation space in regions where additional resolution is needed. This enrichment is introduced through carefully chosen functions that capture local variations and are combined using a partition of unity framework \cite{Babuska1997, Babuska2001}. Within this setting, a solution variable  $u(\bm{x},t)$ is approximated as
\begin{equation}
    u_h( \bm{x}, t) = \sum_{\alpha=1}^n \hat{u}_\alpha(t) \phi_\alpha(\bm{x}) + \sum_{\alpha=1}^n \phi_\alpha(\bm{x}) \sum_{i=1}^m \tilde{u}_{\alpha i}(t) L_i(\bm{x}), \label{gfem}
\end{equation} 
where $u_h(\bm{x}, t)$ denotes the approximate solution at position $\bm{x}$ and time $t$. Here, $n$ is the number of nodes in the domain and $m$ is the number of enrichment functions associated with each node. The functions $\phi_\alpha(\bm{x})$ are the standard Lagrangian finite-element shape functions, while $L_i(\bm{x})$ represent the enrichment functions. The corresponding time-dependent degrees of freedom are given by $\hat{u}_\alpha (t)$ for the conventional FEM component and $\tilde{u}_{\alpha i}(t)$ for the enriched contribution. 
The two terms in Eq. \ref{gfem} play distinct roles. The first corresponds to the standard finite element approximation, while the second augments it with localized enrichment. For each node $\alpha$, this enrichment is supported on a patch formed by the surrounding elements. The corresponding FEM shape function is continuous within this patch and identically zero outside it, which confines the enrichment to a localized region. In contrast, the GFEM shape function
combines the standard shape function $\phi_\alpha(\bm{x})$ with the enrichment function $L_i(\bm{x})$, ensuring compatibility with the finite element framework \cite{Deshmukh2020}.
In the $C^0$-DLS compression approach, the high-fidelity grid is coarsened by grouping high-fidelity nodes into larger, coarser patches, forming GFEM elements. The size of the GFEM domain is specified as $N_I \times N_J \times N_K$, representing the number of nodes along the computational directions $I$, $J$, and $K$, respectively. A single GFEM element, denoted as $(i_n, j_n, k_n)$, is located within the GFEM domain with
$ 1 \leq i_n \leq N_I, \quad 1 \leq j_n \leq N_J, \quad 1 \leq k_n \leq N_K $
Here, the product $N_I \times N_J \times N_K$ equals the total number of GFEM nodes ($n$) as described in equation \ref{gfem}.
The enrichment functions are computed over small, spatially localized sub-domains, referred to as "patches". The size of a patch (in terms of number of high-fidelity points contained) is $(2 \times \text{the size of the GFEM element})^3$ while in discontinuous-DLS the size of the patch is equal to the size of a 3D GFEM element. 
A comprehensive discussion on discontinuous-DLS is provided in the next subsection. Notably, the procedure for learning enrichment functions is the same for both the $C^0$ and discontinuous variants and is also discussed in the following subsection. The spatial localization used in the enrichment function learning process simplifies complex flow structures, enabling the learned basis vectors to more effectively capture high-resolution features \cite{Deshmukh2020, Deshmukh2020a}.
The spatially localized data matrix $[Q]_{S \times M}$, where $S$ represents the number of patches and each patch covers $M$ high-fidelity grid points, is constructed by randomly selecting patches from global high-fidelity observations/snapshots. This approach is commonly used in the field of computer vision. Instead of using the entire training dataset to extract basis modes, the local enrichment functions are learned from a single snapshot of the dataset $Q$ using singular value decomposition (SVD) \cite{Gol-Van-2013}. The training is performed on structured grids within the computational domain $(I, J, K)$ rather than the physical domain $(x, y, z)$.
The $C^0$-DLS approach relies on constructing a global system for data compression using drived enrichment functions. In this method, a set of enrichment functions, coupled with finite element shape functions, is used to compute local element matrices for each element in the dataset. These local matrices are subsequently assembled into a global system, which is then solved to obtain the degrees of freedom matrix, $(S_{R(1+m) \times T}$ , where $R$ is the number of nodes in the GFEM grid, $m$ is the number of enrichment functions per node, and $T$ is the number of global snapshots to be compressed. Data compression is accomplished by storing the basic vectors ($L_i$'s) and the GFEM degrees of freedom ( \( \hat{u}_\alpha\) and \(\tilde{u}_{\alpha i}\)) in a coefficient matrix. A notable strength of the $C^0$-DLS approach is that it preserves $C^0$-continuity in the reconstructed field. Another practical benefit is that the resulting compression ratio is known \emph{a priori}, which provides direct control over the level of data reduction. These benefits, however, come at the cost of having no explicit error bound, which limits the ability to quantify and control reconstruction accuracy.

%
%
%
%
%
%

Therefore, in this work, we extend the $C^0$-DLS methodology by allowing controlled discontinuities across patch boundaries, and shifting the focus toward localized data compression. While the proposed approach retains the underlying ideas of $C^0$-DLS, it replaces global constraints with a more flexible, patch-based treatment of the data and leads to a compression algorithm that works on localized regions of the data. This local perspective allows the method to respond to spatial variation and to balance compression ratio and accuracy in a controlled manner.
%



\subsection{Discontinuous-DLS compression }

This section outlines the main components of the proposed discontinuous-DLS algorithm for error-bounded lossy compression. The algorithm consists of three primary phases: \textit{feature learning}, \textit{compression}, and \textit{decompression}. Each phase plays a distinct role in constructing and applying the compressed representation. The formulation is aimed at compressing large floating-point datasets on structured grids. Since the algorithm operates locally instead of enforcing global assumptions, it can be used for a broad set of datasets and underlying physical systems.
\subsection{Feature learning and compression matrix construction}

The feature learning stage entirely relies on a local enrichment strategy to construct a spatially local basis. This basis is learned from a training dataset or representative snapshot by sampling patches from high-fidelity data. The objective is to capture local spatial structure without requiring access to the full dataset during training. To this end, the data are decomposed into smaller three-dimensional patches of prescribed size. Each patch 
$$\mathbf{P} = \{P_x, P_y, P_z\}, $$
where $P_x$, $P_y$, and $P_z$ denote the number of grid points in the $x$-, $y$-, and $z$-directions, respectively, represents a localized subset of the original field. These patches are selected randomly from the dataset to ensure sufficient variability in the sampled data. The total number of patches, denoted by $S$, is chosen as $4 (P_x \times P_y \times P_z)^3$, which we found to provide a representative sampling of local behavior.
The sampled patches are then arranged into a matrix of size $S \times M$, where each row corresponds to a flattened patch and $M$ is the number of elements within a single patch. This matrix, denoted by ,  $\mathbf{Q}_{S \times M}$ serves as the input for feature extraction and subsequent basis construction.
%
%
Patch size (or coarsening factor) plays an important role in the overall performance of the method. Larger patches generally lead to improved compression ratios but come at a higher computational cost, while smaller patches are less expensive to process but may struggle to represent complex or sparse structures accurately. In practice, the patch size reflects a balance between compression efficiency and reconstruction quality.
%
%
After constructing the sample matrix, singular value decomposition is applied to obtain dominant local features. The SVD decomposes the matrix, $\mathbf{Q}$, as follows:
\begin{equation}
    \mathbf{Q} = \mathbf{U} \mathbf{\Sigma} \mathbf{V}^T,
\end{equation}
%
%
%
where $\mathbf{U} \in \mathbb{R}^{S \times S}$ and $\mathbf{V} \in \mathbb{R}^{M \times M}$ are orthogonal matrices, and $\mathbf{\Sigma} \in \mathbb{R}^{S \times M}$ contains the singular values. The columns of $\mathbf{U}$ form an orthonormal basis for the patch.

In many SVD-based compression methods, only the leading modes are retained, since the singular values often decay rapidly and a low-rank approximation can capture most of the energy in the data \cite{Wang23}. In the present method, however, no truncation is performed. All modes are retained so that the basis spans the full patch space.
%
%
The compression matrix is therefore constructed as
\begin{equation}
    \mathbf{C} = [\bm{\varphi}_1, \bm{\varphi}_2, \dots, \bm{\varphi}_M].
\end{equation}
where $\bm{\varphi}_M$ are the columns of $\mathbf{V}$. Retaining the complete basis ensures that any patch can be represented exactly prior to coefficient approximation, which is necessary to maintain the error-bounded nature of the proposed compression scheme.
The compression matrix $\mathbf{C}$ serves as a localized basis for representing the original data. When constructed using SVD, the resulting basis vectors are orthonormal by definition \cite{Golub2013}, which simplifies both projection and reconstruction. Other decomposition or dictionary-based approaches do not always provide this property. In such cases, additional procedures, such as Orthogonal Matching Pursuit (OMP) \cite{Tropp2004} or Gram–Schmidt orthonormalization \cite{Golub2013}, can be applied to enforce an orthonormal basis.

In error-bound lossy compression, error control is specified at the level of the full dataset through a user-defined global target error $\epsilon_t$. Compression is nevertheless carried out on a patch-by-patch basis, and a corresponding local error tolerance $\epsilon_l$ is introduced to guide the compression within each patch. This local tolerance is derived from the global bound by accounting for the patch size and the number of coarsened elements (or blocks) involved in the approximation. While local error thresholds are used internally during compression, only the global target error is prescribed and enforced.

\subsection{Error Estimation: Local Error Tolerance}

Once the compression matrix has been constructed, reconstruction error is evaluated on each patch to assess compliance with the prescribed global error tolerance. This evaluation provides a quantitative measure of the accuracy of the compressed representation. Two error metrics are considered, depending on how reconstruction fidelity is characterized: the $L^2$ norm error and the $L_\infty$ norm error. Each metric highlights a different aspect of compression accuracy.
The $L^2$ norm is commonly used when a balanced measure of overall error is desired. 
To relate the user-defined global error tolerance $\epsilon_t$ to a local patch-level tolerance, the local error threshold $\epsilon_l$ is computed as follows:
\begin{equation}
   \epsilon_l = \epsilon \cdot \left( \frac{\text{patch size}}{\text{number of coarsened elements}} \right)^{1/2},
\end{equation}
where $\epsilon = \dfrac{ \epsilon_t \|u\| }{100}$ is the global error scaled by the norm of the original data field $u$, and $\|u\|$ denotes the global $L^2$ norm of the snapshot. Here, the ratio of the patch size to the number of coarsened elements accounts for the contribution of each patch to the total error budget.
The absolute $L^2$ error over a patch is defined as:
$$\mathbf{e}_t = \left( \sum_{i=1}^k |u_i - \hat{u}_i|^2 \right)^{1/2},$$
where $u_i$ and $\hat{u}_i$ represent the original and reconstructed fields, respectively, and $k$ is the total number of high-fidelity grid points. 
In this work, the accuracy of the reconstructed data is reported in terms of the relative $L^2$ error, reported as a percentage and referred to here as the normalized root mean square error (NRMSE). It is computed as
$$\text{NRMSE} = 100 \frac{ \left( \sum_{i=1}^k |u_i - \hat{u}_i|^2 \right)^{1/2} }
{ \left( \sum_{i=1}^k |u_i|^2 \right)^{1/2} }.$$

\subsection{Compression methodology}
After learning the features encoded in the compression matrix $\mathbf{C}$ and computing the local error tolerance $\epsilon_l$, compression is applied at the patch level. For each snapshot, a linear system is solved independently for every sampled patch $\mathbf{P}_i^t$, where $i = 1, 2, 3, \dots, N$ denotes the patch index and $t$ is the snapshot index. The system is defined as:
\begin{equation}
  \quad \mathbf{S}^t_i = \mathbf{C}^T \mathbf{P}^t_i.
\end{equation}
Here, $\mathbf{P}_i$ represents the data at the \(i\)-th patch of \(t\)-th snapshot, $\mathbf{C}$ is the compression matrix, and $\mathbf{S}^t_i$ denotes the degrees of freedom (DOFs). These DOFs are sorted in decreasing order of significance, and the optimal set of DOFs is selected by imposing the local error tolerance:
\begin{equation}
   \| \mathbf{C} \tilde{\mathbf{S}}^t_i - \mathbf{P}_i^t \|_2 < \epsilon_l \quad \forall \; i, t.
\end{equation}
Here, the quantity $\mathbf{C} \tilde{\mathbf{S}}_i^t$ denotes the approximate solution obtained by projecting a selected set of degrees of freedom, $\tilde{\mathbf{S}}_i^t$, onto the learned basis $\mathbf{C}$. 
To determine which degrees of freedom (DOFs) should be retained, we progressively test different subsets and examine how well they reconstruct each patch. For every candidate subset, the reconstruction error is evaluated, and the process continues until the prescribed error tolerance is met. To avoid unnecessary trial and error, a bisection approach is used to efficiently narrow down the smallest set of DOFs that satisfies the error criterion.
After this set of DOFs has been selected, the corresponding coefficients are passed through a bit-grooming step. In this stage, the precision of each coefficient is adjusted according to its influence on the reconstructed field. While bit grooming does not reduce data size on its own, it removes insignificant variability in the least significant bits. This cleanup step makes the data much more amenable to subsequent lossless compression. The bit-groomed coefficients are then compressed using the gzip algorithm to further reduce storage requirements. This final compression step preserves the prescribed error bounds while achieving additional data reduction.
Because the method operates on individual patches, $C^0$-continuity  is no longer enforced in the reconstructed field. In practice, relaxing this constraint provides clear gains in flexibility, scalability, and parallel efficiency, which are important for high-performance computing applications. In contrast, the  $C^0$-DLS approach enforces continuity across patches but typically results in larger reconstruction errors for comparable compression ratios.


\begin{algorithm}[t]
\caption{Error-bounded discontinuous-DLS compression algorithm}
\label{alg:eb-dls}
\begin{algorithmic}[1]

\REQUIRE Snapshots $\{\mathbf{u}_1, \mathbf{u}_2, \dots, \mathbf{u}_T\}$, patch size $m$ (with $M = m^3$), target error $\epsilon_t$
\ENSURE Set of sparse coefficient vectors $S_T$ and orthonormal basis vectors $\bm{\Phi}$

\vspace{0.5em}
\STATE \textbf{Step 1: Feature Learning}
\STATE Extract $S=4m^3$ random patches from the initial snapshot $\mathbf{u}_1$ and form the sample matrix $\mathbf{Q} \in \mathbb{R}^{S \times M}$

\STATE Compute SVD of  $\mathbf{Q}$: $\mathbf{Q} = \mathbf{U} \bm{\Sigma} \mathbf{V}^T$
\STATE Set $\bm{\Phi} \gets \mathbf{V}$

\vspace{0.5em}
\STATE \textbf{Step 2: Local Error Tolerance Estimation}
\STATE Partition the domain using coarsening factor $\lambda$
\STATE Compute the local $L_2$ error tolerance $\epsilon_l$ from $\epsilon_t$

\vspace{0.5em}
\STATE \textbf{Step 3: Patch-wise Compression}
\FOR{each snapshot $\mathbf{u}_j$, $j = 1, \dots, T$}
    \FOR{each patch $\mathbf{u}_{j,l}$, $l = 1, \dots, N$}
        \STATE Project onto basis: $\boldsymbol{\alpha}_{j,l} = \bm{\Phi}^T \mathbf{u}_{j,l}$
        \STATE Sort coefficients $\boldsymbol{\alpha}_{j,l}$ in descending magnitude
        \STATE Determine the smallest $n$ such that
        \[
        \left\| \mathbf{u}_{j,l} - \sum_{s=1}^{n} \alpha_{j,l,s} \boldsymbol{\varphi}_s \right\|_{L^2}
        \leq \epsilon_l
        \]
        using a bisection search
        \STATE Retain only the selected coefficients and discard the remainder
        \STATE Apply bit-grooming and GZIP compression to retained coefficients,  $\tilde{\boldsymbol{\alpha}}_{j,l}$
        \STATE Store $\tilde{\boldsymbol{\alpha}}_{j,l}$ in $S_T$
    \ENDFOR
\ENDFOR

\vspace{0.5em}
\RETURN $S_T$, $\bm{\Phi}$

\end{algorithmic}
\end{algorithm}

\subsection{Reconstruction Process for the Discontinuous-DLS Approach}

Reconstruction is carried out in a patch-wise (block-wise) manner. This design allows the procedure to be applied independently across patches and makes parallel execution straightforward. The reconstruction process consists of three main steps: decompression of the retained coefficients, reconstruction of individual patches, and assembly of the full snapshot.

The compressed data are stored as sparse coefficients vectors, also referred to as degrees of freedom (DoFs), $\tilde{\boldsymbol{\alpha}}_{j,l}$, together with the orthonormal basis $\bm{\Phi}$. Before storage, the retained coefficients are processed using bit-grooming followed by GZIP compression. During reconstruction, the compressed coefficients are recovered by applying GZIP decompression and reverse bit-grooming. This yields the coefficient vectors 
$\tilde{\boldsymbol{\alpha}}_{j,l}$, where $j$ denotes the snapshot index and $l$ denotes the patch index. These coefficients correspond to the degrees of freedom associated with each patch.

The data for each patch are reconstructed by forming a linear combination of the basis vectors weighted by the recovered coefficients,
$$\tilde{\mathbf{u}}_{j,l} = \sum_{s=1}^{n} \tilde{\alpha}_{j,l,s}\,\boldsymbol{\varphi}_s, $$
where $\tilde{\mathbf{u}}_{j,l}$ denotes the reconstructed patch data and  $n$ is the number of coefficients or DOFs retained for that patch.

Once all patches corresponding to a given snapshot have been reconstructed, they are assembled to form the complete snapshot,
\begin{equation}
    \tilde{\mathbf{u}_j} = \bigcup_{l=1}^{N} \tilde{\mathbf{u}}_{j,l},
\end{equation}
where $\mathbf{u}_j$ represents the reconstructed snapshot at index $j$, and $\bigcup$ denotes the assembly of the individual reconstructed patches into the full field. This procedure is repeated for all snapshots to recover the complete time-varying dataset. Owing to its patch-wise and element-wise structure, the reconstruction process is highly parallelizable and scales well to large scientific datasets.

\begin{algorithm}[t]
\caption{Decompression algorithm}
\label{alg:decompression}
\begin{algorithmic}[1]

\REQUIRE Sparse coefficient vectors $\{\tilde{\boldsymbol{\alpha}}_{j,l}\}$ and orthonormal basis $\bm{\Phi} = [\bm{\varphi}_s]_{s=1}^{M}$

\ENSURE Reconstructed snapshots $\{\mathbf{u}_1, \mathbf{u}_2, \dots, \mathbf{u}_T\}$

\vspace{0.5em}
\STATE Recover the decompressed coefficients $\{\tilde{\alpha}_{j,l}^s\}$ using GZIP decompression and reverse bit-grooming

\FOR{each snapshot $j = 1, \dots, T$}
    \STATE Initialize reconstructed snapshot $\mathbf{u}_j$
    \FOR{each patch $l = 1, \dots, N$}
        \STATE Reconstruct patch $l$ using:
        \STATE \quad $\mathbf{u}_{j,l} = \sum_{s=1}^{n_{j,l}} \tilde{\alpha}_{j,l}^s \bm{\varphi}_s$
    \ENDFOR
    \STATE Concatenate patches to form full snapshot:
    \STATE \quad $\mathbf{u}_j = \bigcup_{l=1}^{N} \mathbf{u}_{j,l}$
\ENDFOR

\vspace{0.5em}
\RETURN $\{\mathbf{u}_1, \mathbf{u}_2, \dots, \mathbf{u}_T\}$

\end{algorithmic}
\end{algorithm}

\subsection{Implementation}

The $C^0$-DLS and discontinuous-DLS algorithms are implemented in a C++ library leveraging PETSc \cite{petsc, petscsf2022} for linear algebra operations and domain decomposition-based parallelization.

\subsubsection{Feature Learning Phase}

The feature learning phase constructs a distributed $S \times M$ PETSc matrix, where $N$ represents the number of nodes in a patch and $S$ is the number of required samples ($S \geq M$). The samples (i.e, matrix rows) are distributed across processors according to PETSc's default domain decomposition strategy:

\begin{align}
\text{Div} &= \lfloor \text{Total\_Rows} / \text{Total\_Procs} \rfloor \\
\text{Rem} &= \text{Total\_Rows} \bmod \text{Total\_Procs} \\
\text{rows\_on\_proc}_p &= \text{Div} + (p < \text{Rem} ? 1 : 0) \\
\text{row\_start\_proc}_p &= \text{Div} \times p + \min(p, \text{Rem})
\end{align}

Following matrix assembly, parallel singular value decomposition (SVD) is performed using the cross-product based method implemented in SLEPc \cite{slepc}. An MPI gather operation is used to form a local copy of the learned SVD basis (i.e, the right singular vectors of $S$) on each process.   

\subsubsection{Compression and Decompression}

For compression, each processor compresses a predetermined number of patches using the same distribution scheme as feature learning. To optimize I/O performance, patches are processed in batches of 1000, with \texttt{MPI\_Scan} operations determining write positions for variable-length compressed DOF arrays. 

Compressed data is stored in a binary format using MPI-IO. The binary format includes a fixed size, addressable header that contains information and the size and location of the compressed DOF array for each patch in the file. The compressed DOF arrays are stored in a tightly packed binary array that starts directly after the header. When available, multithreading is used to minimize the effect of I/O costs on total runtime.

Decompression follows an analogous parallel strategy where each processor handles a fixed subset of patches. Decompressed values are populated into a global reconstruction vector, requiring minimal inter-processor communication for final assembly.


\section{Comparison with the state-of-the-art lossy compression techniques}

The dataset under consideration in this study is derived from a three-dimensional incompressible CFD simulation of turbulent flow past a stationary cylinder, generated using implicit large eddy simulation (LES) at a Reynolds number of 100,000. The computational domain employs a curvilinear grid of size 695 × 396 × 149, surrounding a rigid cylinder of unit diameter, with a radius of 97.2 units and a spanwise extent of 0.4 units. The simulation applies uniform freestream conditions at the inflow and transverse boundaries, no-slip and isothermal conditions at the cylinder surface, and periodic boundary conditions in the spanwise direction. The full dataset consists of 3000 snapshots, but for the demonstration of the developed discontinuous-DLS compression algorithm, a representative subset of 1024 snapshots was selected. These snapshots capture statistically stationary fluctuating velocity fields over time, resulting in approximately 937.5 GB of high-fidelity data \cite{Deshmukh2020a}. This subset preserves the essential unsteady flow dynamics, including vortex shedding and turbulent wake structures, making it a suitable target for evaluating data compression methods. 
All compression and decompression experiments were carried out on the Stokes HPC cluster at the University of Central Florida, USA.

\begin{figure}[h]
    \centering
    \includegraphics[width=\linewidth]{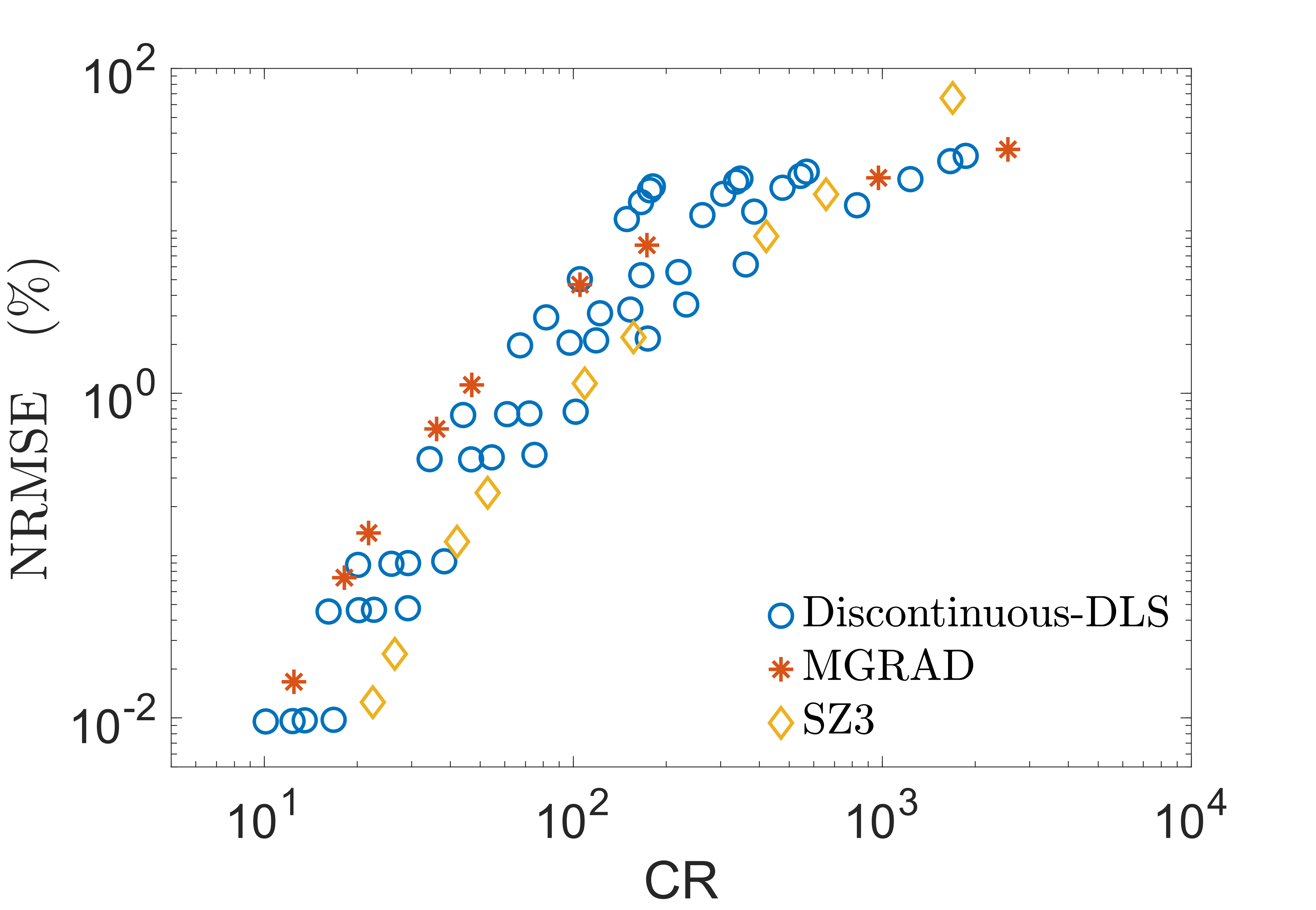}
    \caption{The error bound discontinuous-DLS based comporession results is compared to the SZ, MGARD, and continuous-DLS based compression of the 3D turbulent flow.}
    \label{fig:comparision}
\end{figure}

To evaluate the performance of the proposed error-bound discontinuous-DLS compression technique, we first compare it with two benchmark error-bounded lossy compressors--MGARD \cite{Ainsworth2020, Qian23} and SZ3 \cite{Liang23}, using normalized root mean square error (NRMSE) and compression ratio (CR) as performance metrics. Results for all three methods, SZ3, MGARD, and the proposed discontinuous-DLS, are summarized in figure \ref{fig:comparision}.

MGARD is a state-of-the-art multilevel lossy compressor developed for datasets defined on structured and unstructured grids. In this study, we use its CPU-based implementation, which performs 
a multilevel decomposition of the input data, treating it as a piecewise multilinear function and successively constructing coarser representations through interpolation. At each level of the hierarchy, MGARD computes and stores an interpolated approximation of the data using a coarser grid and subtracts this approximation from the original data to generate multilevel coefficients (residuals), which capture the high-frequency details lost in the interpolation. These coefficients are then projected onto the coarser grid to estimate a correction term, which is added back to the coarse-level data to improve the approximation. This process continues recursively until the coarsest level is reached, where the final coarse representation is stored directly. These multilevel coefficients and the final coarse representation are then compressed using Huffman coding and other lossless techniques. 
%
%

SZ3 is a lossy compression algorithm that uses a prediction-based approach to compress scientific data. It predicts the values of the data based on the surrounding values and encodes the residuals (the differences between the predicted and actual values). 
The effectiveness of this approach stems from its ability to exploit local smoothness in the data. In practice, SZ3 relies on piecewise linear regression for prediction and combines this with entropy coding to reduce storage cost. 
%

Notably, MGARD and SZ3 are not natively designed to operate under NRMSE-based error bounds; they typically support absolute or relative error constraints. The NRMSE values reported here were computed retrospectively from the decompressed data with respect to the original dataset, ensuring a fair and consistent evaluation across all three methods.
The comparative results in Figure \ref{fig:comparision} show that the proposed DLS-based approach exhibits both tunability and scalability. Across the tested range, compression ratios span from $10$x to over $1800$x as the NRMSE increases from approximately $\sim 0.001$ to nearly $25$. 
%
%
In our experiments, MGARD achieves a compression ratio of about $15\times$ at $\text{NRMSE} \approx 0.0125$,  increasing to more than $2500\times$ at $\text{NRMSE} \approx 25$. At lower error levels, however, MGARD consistently underperforms discontinuous DLS, often yielding compression ratios more than 50\% lower. This gap gradually diminishes as the error tolerance increases. SZ3 shows a different trend: it performs very well at low error levels (for example, reaching CR $\sim 22$x at NRMSE $\sim 0.0125$), but becomes comparable to the proposed discontinuous-DLS at more relaxed error bounds. At moderate to high error levels, discontinuous-DLS outperforms both MGARD and SZ3.

Beyond error-compression behavior, the proposed approach also differ in how temporal information is treated. MGARD and SZ3 operate on each snapshot independently and do not attempt to exploit correlations across time. In contrast, discontinuous DLS constructs a compression matrix from the first snapshot and reuses it for subsequent snapshots, allowing temporal coherence to be leveraged. To maintain a fair comparison, all 1024 snapshots were compressed independently using MGARD, SZ3, and current approach. We observed that the overall compression ratios may differ by approximately $\pm$5–10\% as a result of this processing strategy.

While $C^0$-DLS is capable of achieving extremely high compression ratios (up to 2570x)  \cite{Deshmukh2020a}, this method offers limited flexibility and is not well suited for applications where reconstruction accuracy must be controlled. Discontinuous-DLS, by contrast, provides a more balanced alternative. It occupies a middle ground between error-bounded compressors like SZ3 and aggressively compressive approaches like $C^0$-DLS, offering a controllable trade-off between accuracy and compression efficiency.
Overall, the comparison results suggest that discontinuous-DLS achieves a favorable balance between compression efficiency and reconstruction accuracy. The method exhibits adaptive behavior across error levels and consistently attains moderate-to-high compression ratios while maintaining error control. These observations provide useful context for the more detailed analyses presented in the following sections.

In the discussion that follows, we report results in terms of the coarsening factor rather than the patch size. The coarsening factor is computed based on the patch size and is defined as the ratio of the total number of high-fidelity grid points to the number of coarsened grid points determined by the given patch size.

\section{Role of basis selection in data-driven compression}

\begin{figure}
    \centering
    \includegraphics[width=\linewidth]{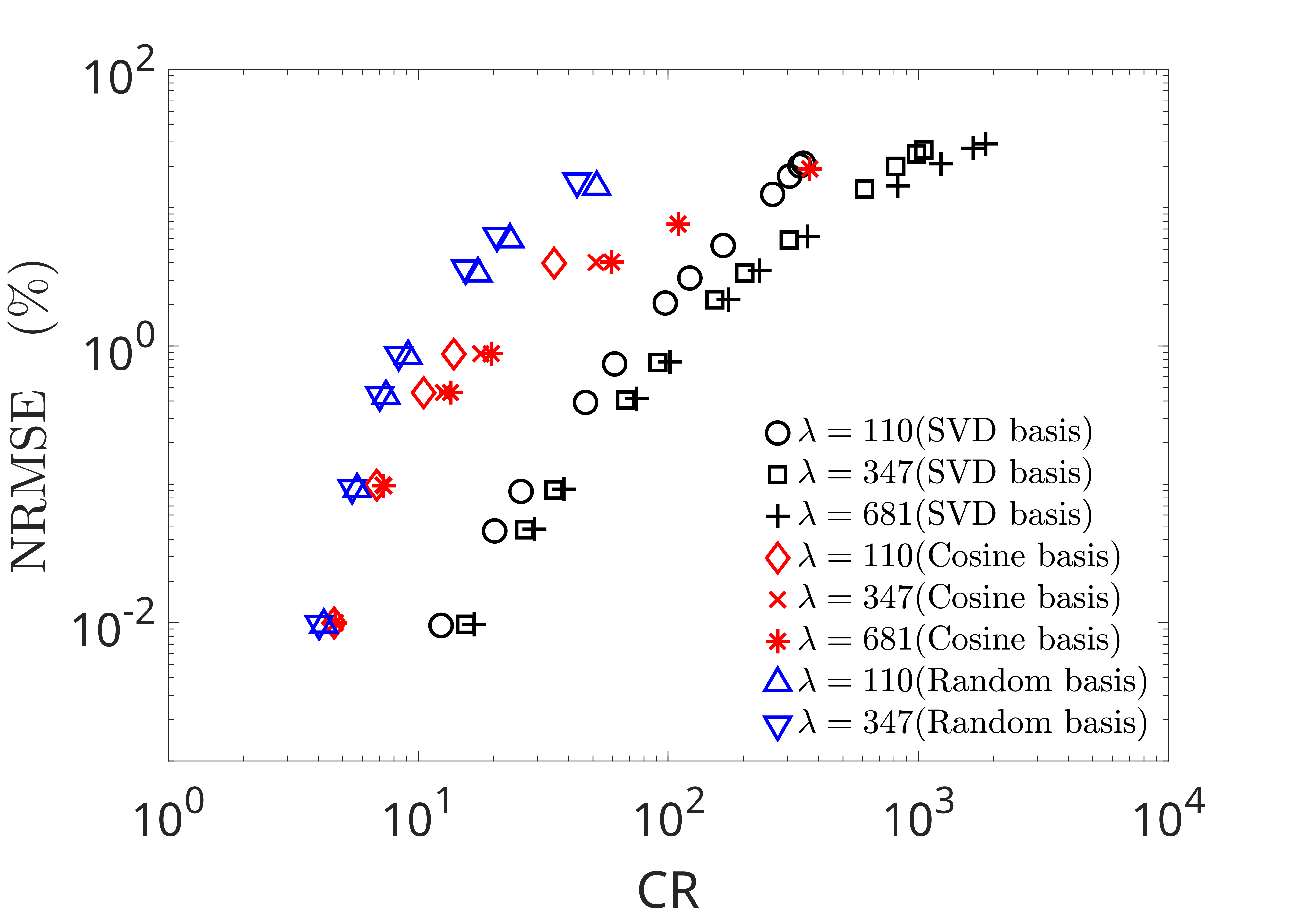}
    \caption{Comparison of compression performance (CR versus NRMSE) using three types of basis functions: data-adaptive SVD, fixed cosine, and random bases. } 
    \label{fig:comparision_basis}
\end{figure}

This section explores how the choice of basis influences the performance of the proposed compression approach. In particular, we compare SVD, cosine, and random bases. Figure~\ref{fig:comparision_basis} summarizes this comparison by plotting compression ratio against normalized root mean square error for different coarsening factors.

Among the three options, the SVD basis consistently offers the most favorable balance between compression ratio and reconstruction accuracy. This trend can be attributed to the data-driven nature of the SVD modes, which are able to adapt to the dominant features present in the data. Even under more aggressive compression, essential structures are largely preserved, with reconstruction error increasing in a gradual and predictable manner as the coarsening factor grows.

The cosine basis shows moderate compression performance along with relatively stable reconstruction accuracy, benefiting from its fixed and computationally efficient modes. However, its lack of adaptivity limits its ability to represent data-specific features. In contrast, the random basis performs poorly across all coarsening factors, with reconstruction error increasing rapidly as the compression ratio rises.

Taken together, these results highlight he importance of basis selection in data-driven compression and provide a clear motivation for using SVD-based adaptive modes in the remainder of this work.

\section{Compression analysis on a single snapshot}

\begin{figure*}[htbp]
    \centering
    \begin{subfigure}[b]{0.45\linewidth}
        \centering
        \includegraphics[width=\linewidth]{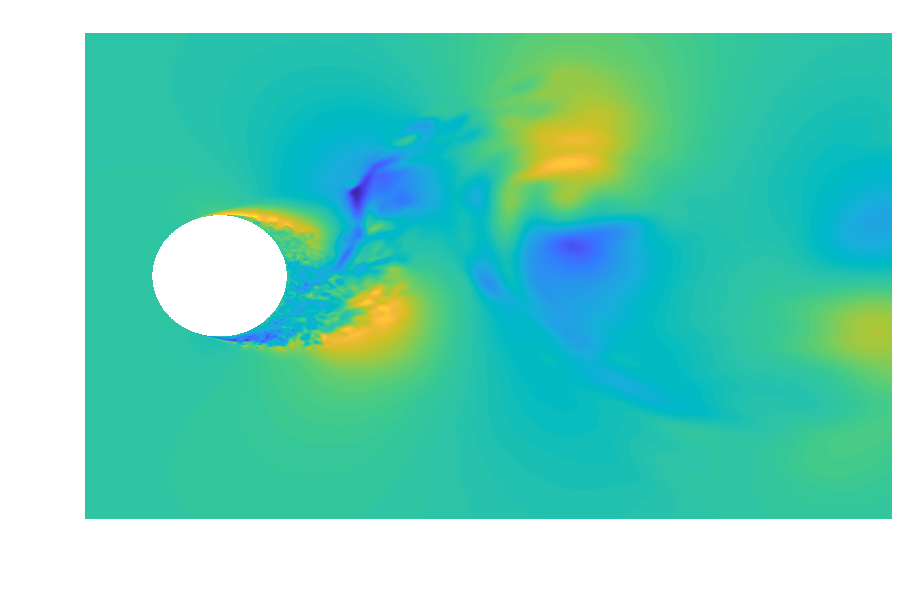}
        \caption{2D $xy$ slice of the training snapshot}
        \label{fig:train}
    \end{subfigure}
    \begin{subfigure}[b]{0.45\linewidth}
        \centering
        \includegraphics[width=\linewidth]{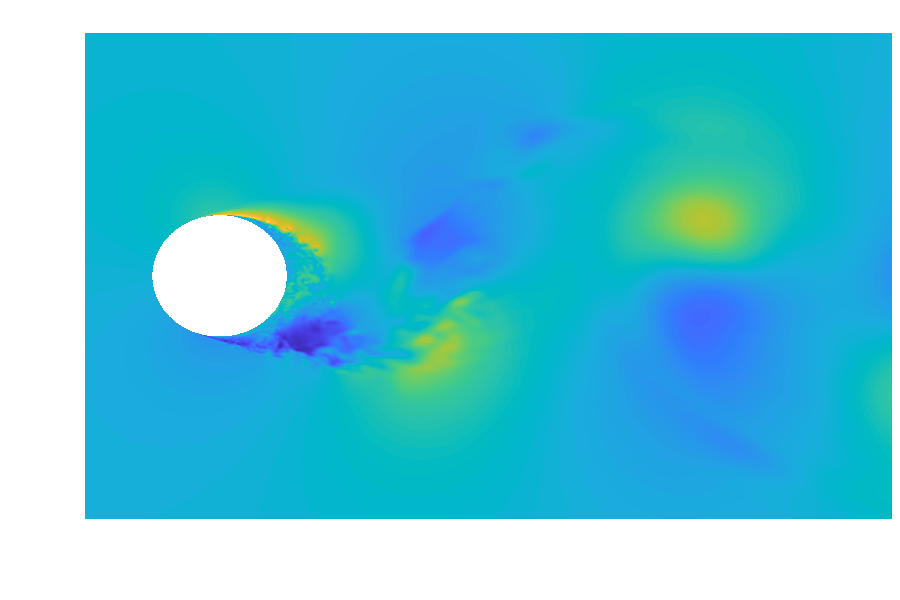}
        \caption{2D $xy$ slice of the test snapshot}
        \label{fig:test}
    \end{subfigure}
    \caption{The training snapshot is used to learn the enrichment functions, while the test snapshot is utilized for parametric study. Although only the $u'$ component is displayed here, all three velocity components ($u'$, $v'$, and $w'$) are included in both training and testing.}
    \label{fig:train-test-snap}
\end{figure*}

Having established the role of basis selection, we now examine the compression behavior of the proposed method in more detail. The discussion starts with a single snapshot in this section and then moves to the full time-series dataset of 1024 snapshots in Section~6. In all cases, the compression matrix $\mathbf{C}$ is built using an SVD basis learned from a single three-dimensional training snapshot of the fluctuating velocity field, as shown in Fig.~\ref{fig:train-test-snap}(a).
Separate basis sets are constructed for each velocity component (\(u'\), \(v'\), and \(w'\)), allowing the method to adapt component-wise across different coarsening factors or patch sizes. Throughout the following discussion, results are reported in terms of the coarsening factor rather than the patch size, as defined earlier.

To examine how the method performs over time, a different snapshot from the time-series dataset is selected for testing, as shown in Fig.~\ref{fig:train-test-snap}(b). Compression performance and reconstruction accuracy are evaluated using the compression ratio and normalized root mean square error (NRMSE). A parametric study is carried out by varying the coarsening factor and the target error bound. In addition, the vorticity magnitude is examined to assess how well key flow features are preserved in the reconstructed fields.

\begin{figure}[h!]
    \centering
        \includegraphics[width=\linewidth]{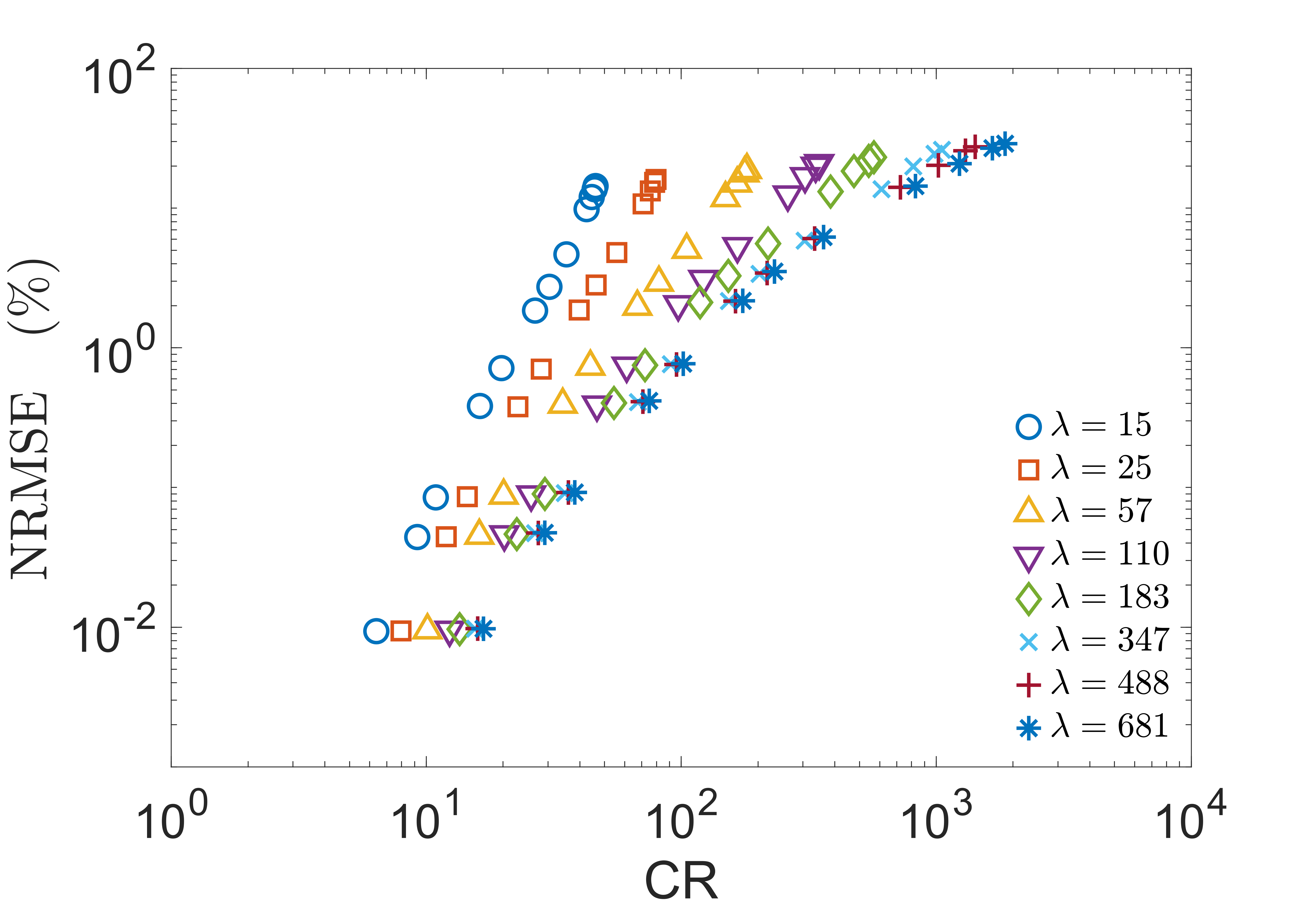}
    \caption{Normalized root mean square errors (NRMSEs) and compression ratios (CRs) for 1024 snapshots of flow past a cylinder, computed using various patch sizes and target error bounds.}
   \label{fig:NRMSE_CR}
\end{figure}

Figure \ref{fig:NRMSE_CR} shows how the compression ratio varies with normalized root mean square error. Results are reported for a range of  \textit{coarsening factors} ($\lambda$) spanning from 15 to 681, which correspond to patch sizes between 5 and 19. As expected, an inverse relationship between compression ratio and reconstruction error is observed: achieving lower NRMSE generally requires retaining more information and therefore results in reduced compression. For a fixed error level, larger coarsening factors consistently yield higher compression ratios, indicating that increasing the coarsening factor improves overall compression efficiency. For instance, at an NRMSE of approximately 0.0095\%, the compression ratio increases from about $6 \times$ for a coarsening factor of 15 to more than $19\times$ for a coarsening factor of 681. This trend becomes increasingly pronounced as the prescribed error is relaxed. At moderate error levels (NRMSE ~2–3\%), coarsening factors of 110 and higher achieve compression ratios exceeding $100 \times$, with $\lambda=183$ reaching roughly $569 \times$  and $\lambda=681$ approaching  $1860 \times$.

The improved compression at larger coarsening factors can be attributed to their ability to capture and exploit large-scale spatial correlations within the data, thereby reducing redundancy more effectively. At the same time, the results also indicate diminishing returns beyond a coarsening factor of approximately 347, where further increases in $\lambda$ lead to only marginal gains in compression ratio. This behavior is largely due to the growing overhead associated with basis storage and a gradual loss of sparsity in the retained coefficients.

Another noteworthy observation is that the achieved NRMSE values are consistently about an order of magnitude lower than the user-specified target error bounds, particularly at larger coarsening factors. This conservative behavior, where the actual reconstruction error remains below the prescribed tolerance, has been reported for other established lossy compression methods \cite{Lakshminarasimhan2013, Cappello2019, Di2016}. 
Overall, the results demonstrate that discontinuous DLS provides effective, error-controllable, and high-fidelity compression, with moderate to large coarsening factors (approximately 110–183) offering a particularly favorable balance between compression efficiency and reconstruction accuracy for large-scale, complex datasets.


\begin{figure}[h!]
    \centering
        \includegraphics[width=\linewidth]{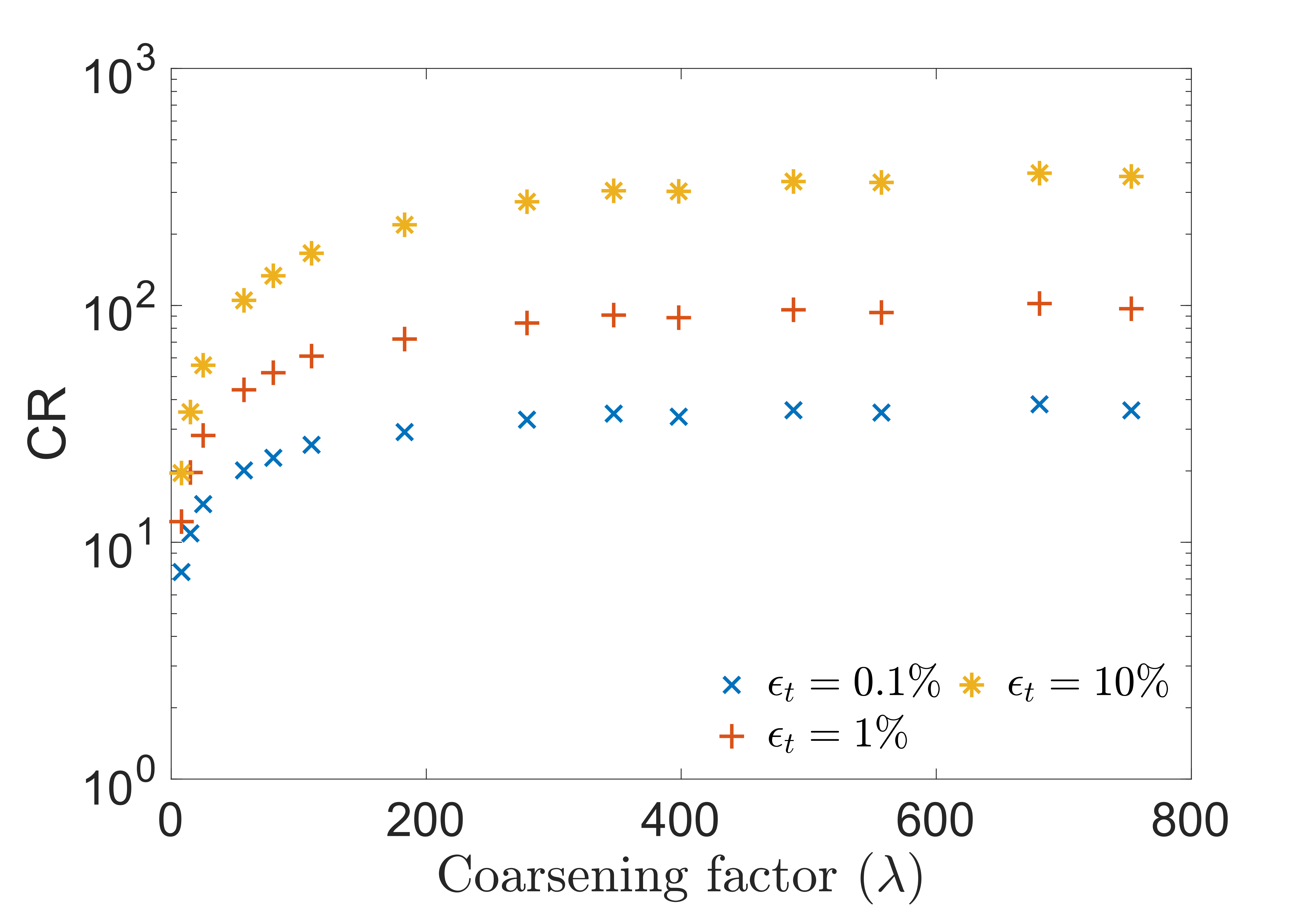}
    \caption{Estimated compression ratios for varying patch sizes at three target error levels: 0.1\%, 1\%, and 10\%. The estimates are based on compression ratios computed from a single snapshot and extrapolated to 1024 snapshots. }
    \label{fig:CR_vs_patch}
\end{figure}

Next, we examine how compression efficiency varies with patch size, with the goal of identifying an optimal choice for practical applications. Figure \ref{fig:CR_vs_patch} shows how the compression ratio, computed over all 1024 snapshots, changes with the coarsening factor for three target error levels: 0.1\%, 1\%, and 10\%. Across all target errors, a similar trend is observed. As the coarsening factor increases, the compression ratio rises steadily and continues to improve up to a coarsening factor of approximately 300. Beyond this point, the curves begin to level off, indicating that further increases in coarsening factor yield only marginal gains in compression efficiency. The maximum compression ratios achieved are approximately $30\times$ for a target error of 0.1\%, $55\times$ for 1\%, and $290\times$ for 10\%. These results suggest that moderate coarsening factors, typically in the range of 10 to 200, offer a practical balance between preserving reconstruction accuracy and achieving meaningful compression. Although larger coarsening factors (around 200 and above) can produce higher compression ratios, the incremental benefits become increasingly limited. Therefore, the choice of coarsening factor should be guided by the desired trade-off between compression efficiency and acceptable reconstruction error, particularly when working with large-scale or time-dependent datasets.

\begin{figure*}[htbp]
    \centering

    \begin{subfigure}[b]{0.45\textwidth}
        \includegraphics[width=\linewidth]{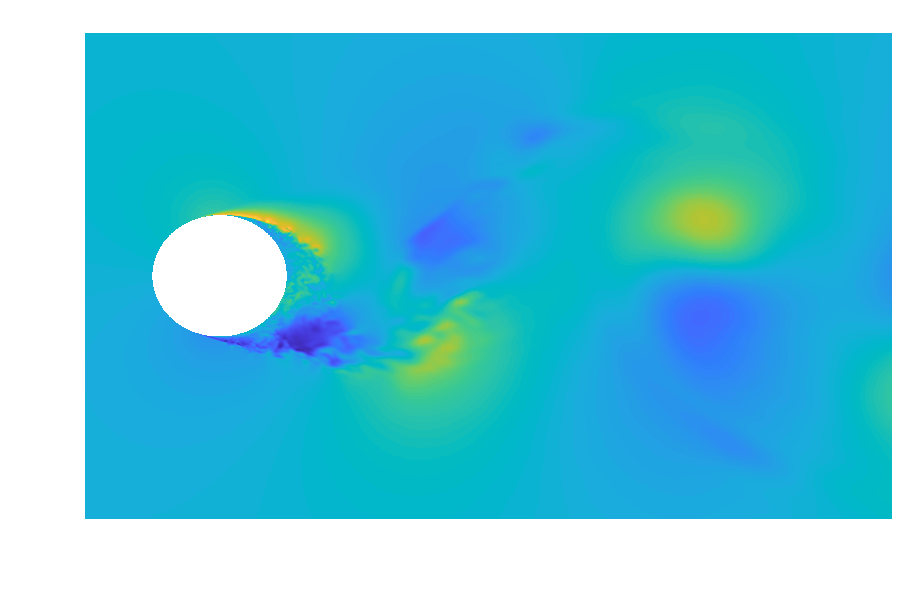}
        \caption{Test snapshot}
        \label{fig:A}
    \end{subfigure}
    \begin{subfigure}[b]{0.45\textwidth}
        \includegraphics[width=\linewidth]{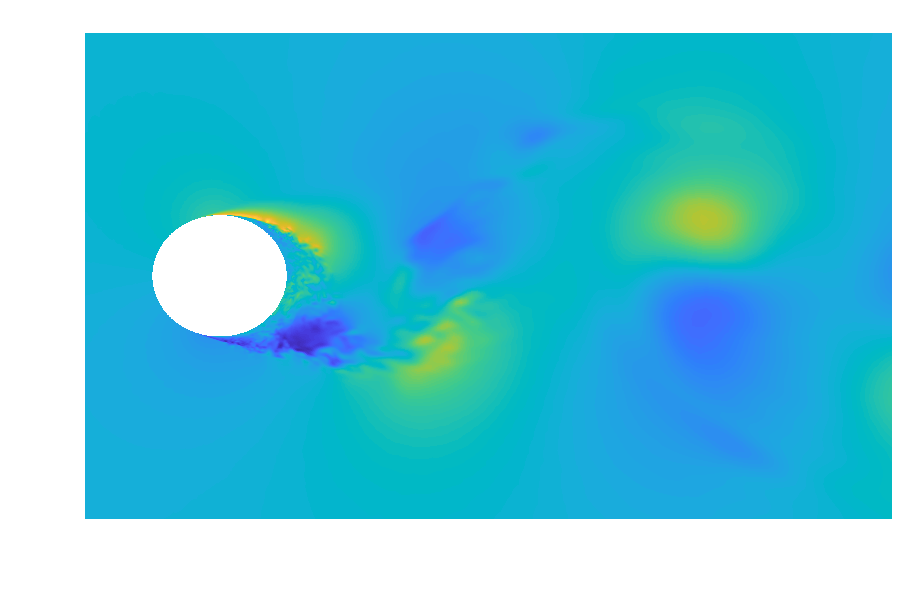}
        \caption{NRMSE: 0.391\%, CR: 47$\times$}
        \label{fig:B}
    \end{subfigure}

    \vspace{1ex}

    \begin{subfigure}[b]{0.45\textwidth}
        \includegraphics[width=\linewidth]{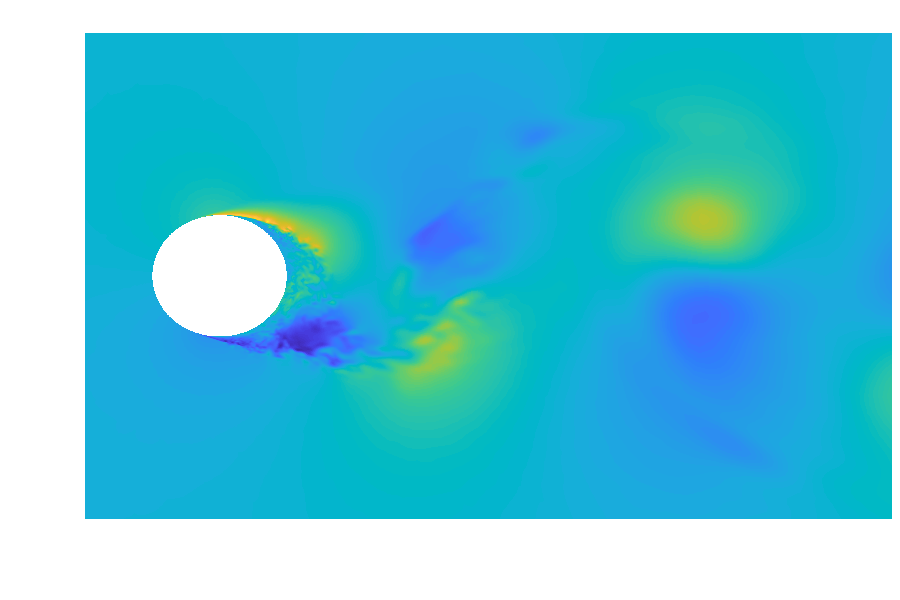}
        \caption{NRMSE: 0.408\%, CR: 68$\times$}
        \label{fig:C}
    \end{subfigure}
    \begin{subfigure}[b]{0.45\textwidth}
        \includegraphics[width=\linewidth]{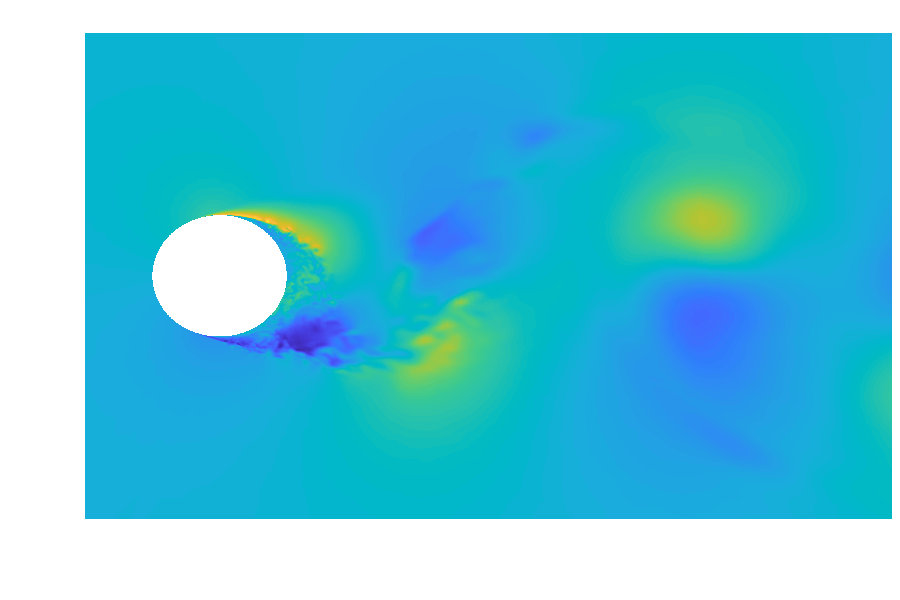}
        \caption{NRMSE: 0.8\%, CR: 91$\times$}
        \label{fig:D}
    \end{subfigure}

    \vspace{1ex}

    \begin{subfigure}[b]{0.45\textwidth}
        \includegraphics[width=\linewidth]{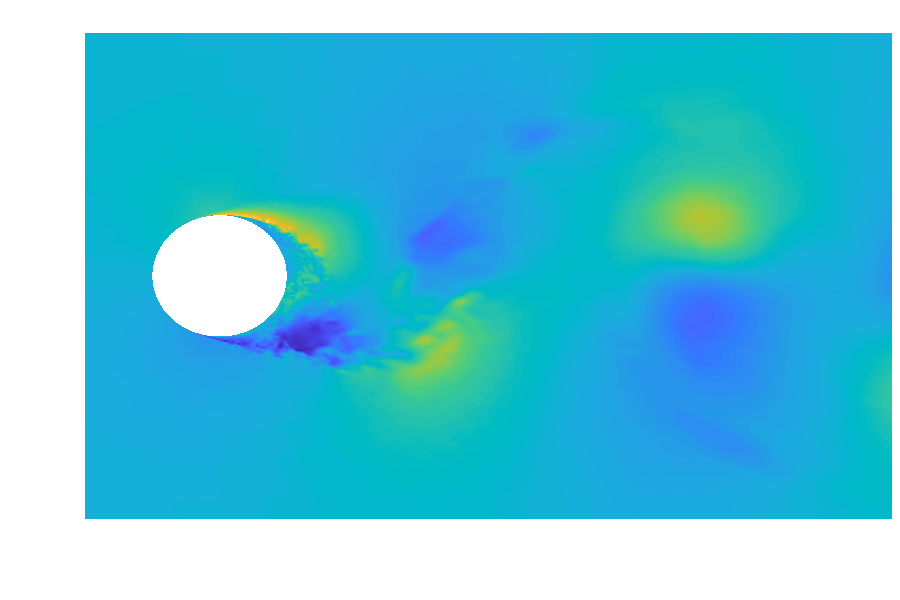}
        \caption{NRMSE: 3.10\%, CR: 122$\times$}
        \label{fig:E}
    \end{subfigure}
    \begin{subfigure}[b]{0.45\textwidth}
        \includegraphics[width=\linewidth]{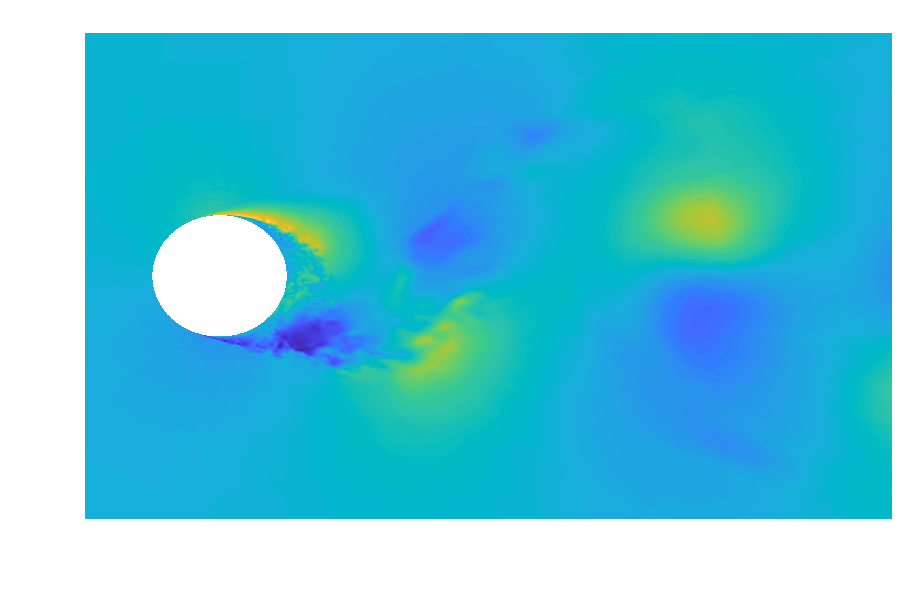}
        \caption{NRMSE: 3.43\%, CR: 217$\times$}
        \label{fig:F}
    \end{subfigure}

    \caption{
        Zoomed-in 2D slices of the \( u' \) component from the 3D flow past a cylinder, comparing the test snapshot with various discontinuous-based reconstructions.
    }
    \label{fig:velocity}
\end{figure*}

\begin{figure*}[h!]
    \centering
  \centering
    \begin{subfigure}[b]{0.45\linewidth}
        \centering
         \includegraphics[width=\linewidth]{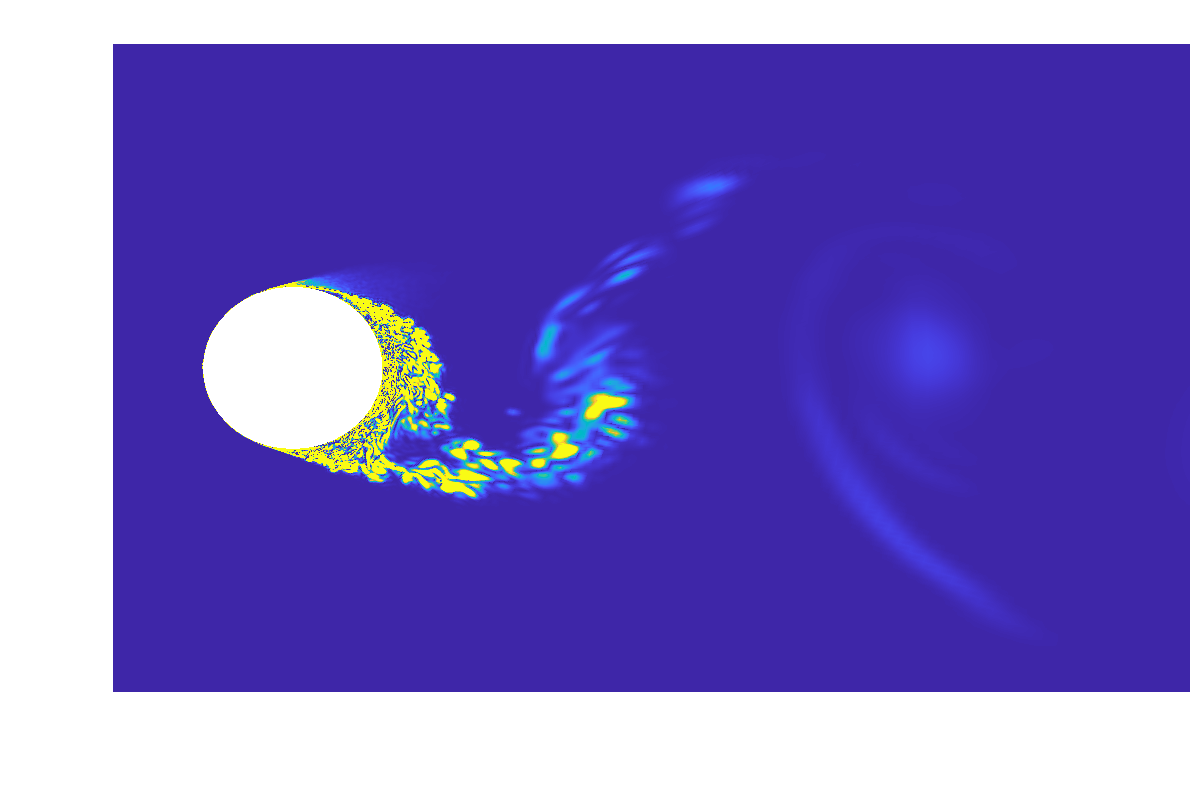}
                         \vspace{-0.3in}
        \caption*{(A) Test snapshot}
    \end{subfigure}
    \begin{subfigure}[b]{0.45\linewidth}
        \centering
          \includegraphics[width=\linewidth]{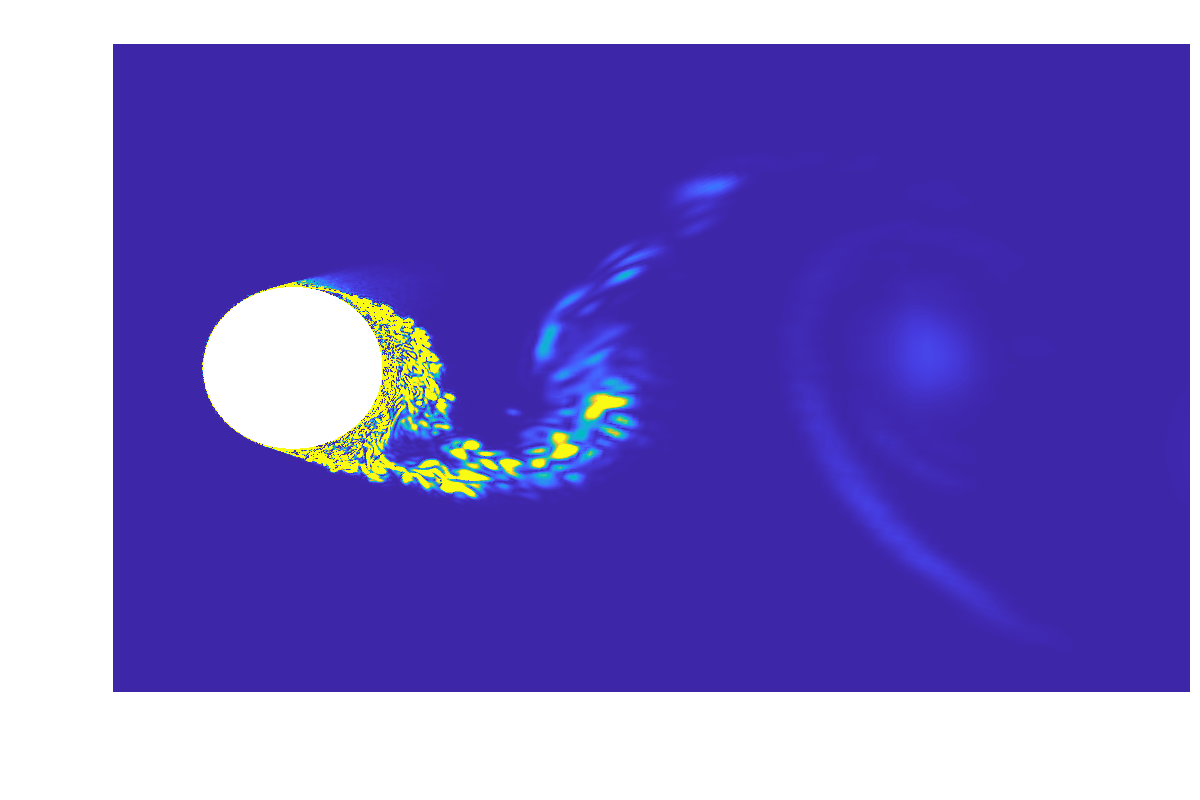}
                 \vspace{-0.3in}
        \caption*{(B) NRMSE: 5.41\%, CR: 47$\times$}
    \end{subfigure}


  \centering
    \begin{subfigure}[b]{0.45\linewidth}
        \centering
        \includegraphics[width=\linewidth]{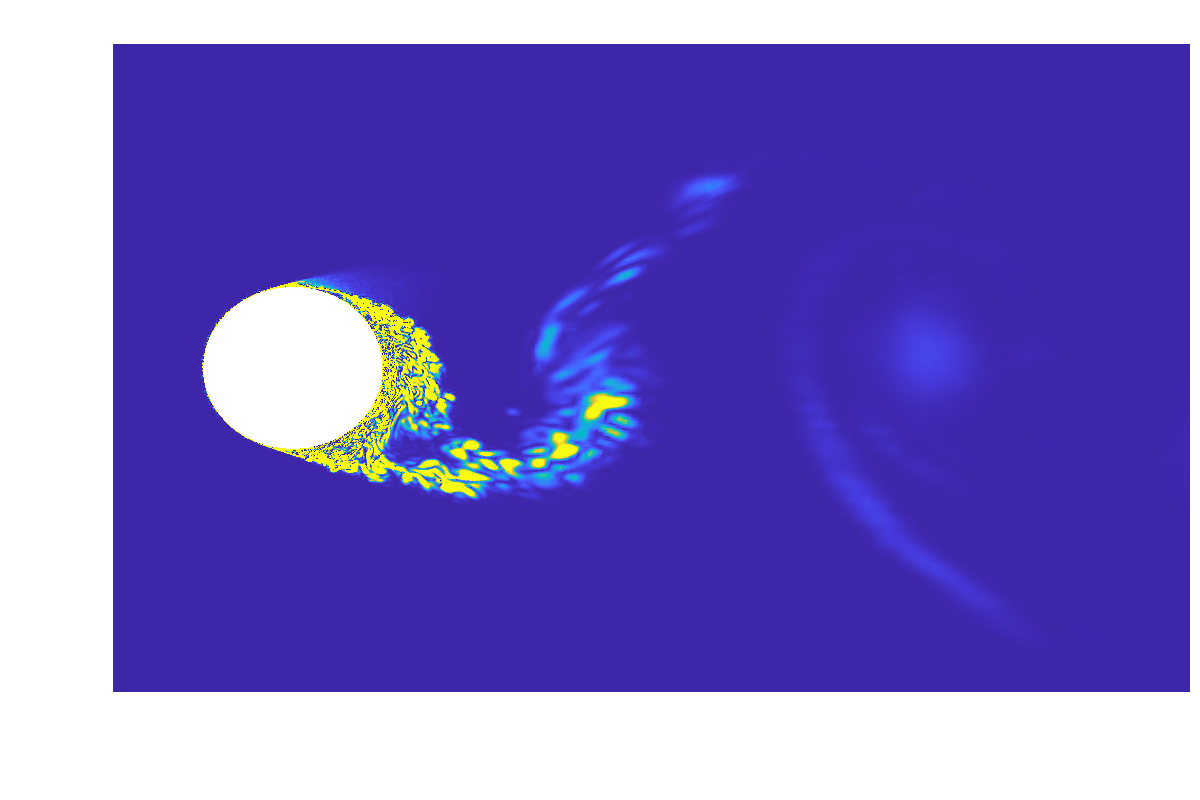}
                 \vspace{-0.3in}
        \caption*{(C) NRMSE: 5.36\%, CR: 68$\times$}
    \end{subfigure}
    \begin{subfigure}[b]{0.45\linewidth}
        \centering
         \includegraphics[width=\linewidth]{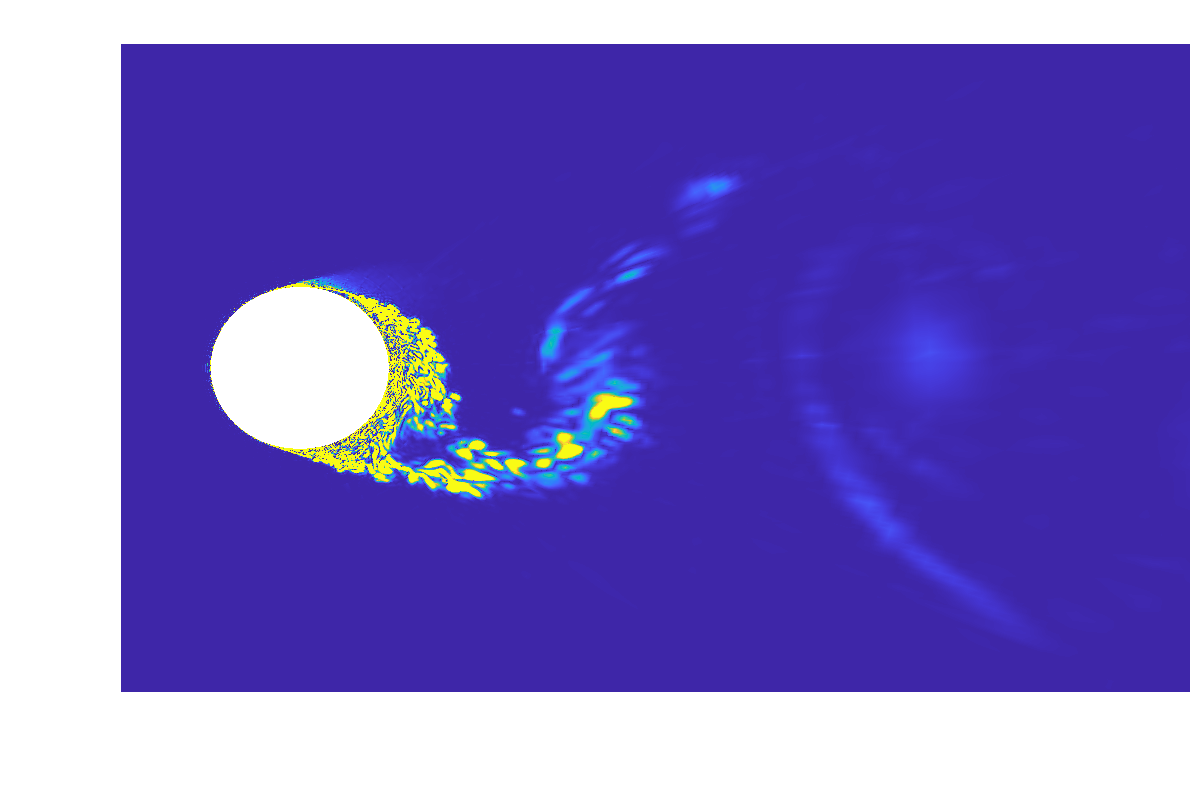}
                 \vspace{-0.3in}
        \caption*{(D) NRMSE: 9.72\%, CR: 91$\times$}
    \end{subfigure}


  \centering
    \begin{subfigure}[b]{0.45\linewidth}
        \centering
         \includegraphics[width=\linewidth]{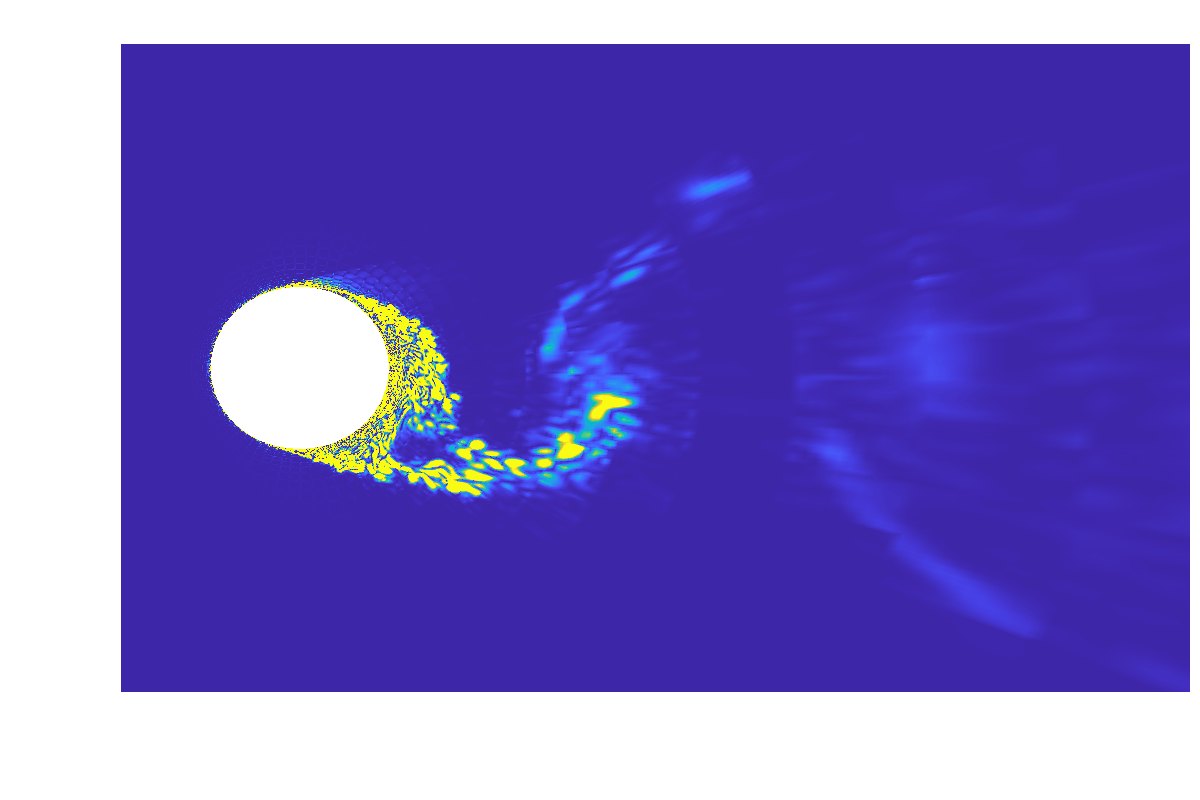}
                 \vspace{-0.3in}
        \caption*{(E) NRMSE: 41.56\%, CR: 122$\times$}
    \end{subfigure}
    \begin{subfigure}[b]{0.45\linewidth}
        \centering
         \includegraphics[width=\linewidth]{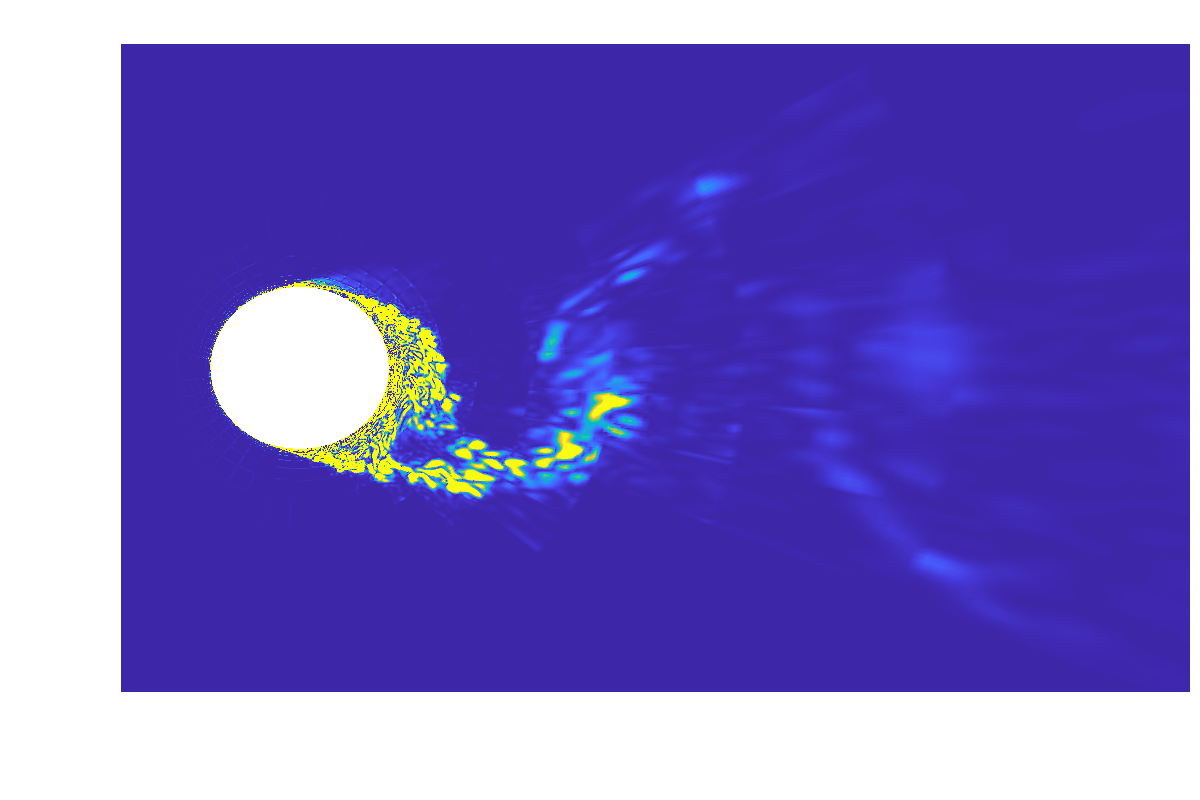}
                 \vspace{-0.3in}
        \caption*{(F) NRMSE: 35.97\%, CR: 217$\times$}
    \end{subfigure}
    \caption{Zoomed-in 2D slices of the vorticity magnitude from various reconstructions using the discontinuous-DLS method, compared against the test snapshot.  (A) Test snapshot, (B) Target error=0.5\% and $\lambda=110$, (C) Target error=0.5\% and $\lambda=347$, (D) Target error=1\%  and $\lambda=347$, (E) Target error=5\% and $\lambda=110$, and (F) Target error=5\% and $\lambda=488$.}
        \label{fig:vorticity}
\end{figure*}

To assess the quality of the reconstructed flow fields produced by the discontinuous-DLS decompression, we examine instantaneous snapshots of the velocity and the calculated vorticity magnitude computed from them. Five representative cases are selected to asses a broad range of coarsening factors and target error tolerances: $(110,,0.5\%)$, $(347,,0.5\%)$, $(347,,1\%)$, $(110,,5\%)$, and $(488,,5\%)$. Together, these cases provide a clear illustration of how reconstruction fidelity varies across different compression settings. The actual reconstruction errors corresponding to these cases are $0.391\%$, $0.408\%$, $0.8\%$, $3.10\%$, and $3.43\%$, respectively. The reconstructed fields are visually compared with the original solution, with figure \ref{fig:velocity} showing the velocity fields and figure \ref{fig:vorticity} illustrating the corresponding vorticity magnitude.

At the lower target error of 0.5\%, the reconstructed velocity fields remain in close agreement with the original solution, as illustrated in figure \ref{fig:velocity}(A–C). The associated vorticity magnitude fields also exhibit smooth spatial variations, with no visible artifacts or discontinuities (see figure \ref{fig:vorticity}(A–C)). Despite the lack of enforced $C^0$-continuity in the discontinuous DLS framework, the reconstructions for these low-error cases show no apparent artificial discontinuities, even in regions with strong shear or vortex shedding. These results indicate that when the compression error is tightly controlled, the reconstructed variables, and consequently derived quantities such as vorticity remain both physically meaningful and accurate. Under these conditions, compression ratios of up to $68\times$ can be achieved, highlighting the ability of the method to significantly reduce data size without compromising essential flow characteristics.

When the target error level increases to 1\%, a gradual loss of fine-scale detail becomes noticeable. For the case $(347,,1\%)$, which corresponds to an actual error of $0.8\%$ (see figure \ref{fig:velocity}(D)), the reconstructed velocity field remains largely consistent with the original solution. However, slight smoothing of sharp gradients and a modest reduction in vorticity intensity can be observed. Despite these localized effects, the overall flow organization--including the vortex structure and wake dynamics--remains well preserved  (see figures \ref{fig:vorticity}(A) and \ref{fig:vorticity}(D)). Notably, this case attains a compression ratio of up to $91\times$, demonstrating that substantial data reduction is still possible while retaining the dominant flow features.

More noticeable differences appear in the higher target error levels, $(110,,5\%)$ and $(488,,5\%)$, where the reconstruction error exceeds 3\%. In these cases, the velocity fields show mild smoothing in regions with strong gradients, and the corresponding vorticity magnitude fields display a visible reduction in fine-scale detail. Despite this loss of resolution, the vorticity magnitude in the immediate vicinity of the cylinder remains largely intact, preserving the dominant vortex structures that characterize the near-wake region. Further downstream, however, clearer visual differences emerge as smaller-scale vortical features become increasingly diffused. This behavior suggests that while the primary flow dynamics are retained, the finer structures in the far wake are more sensitive to higher levels of lossy compression. Importantly, even under these more aggressive compression settings, achieving compression ratios of up to approximately $220\times$, the reconstructed fields do not exhibit any discontinuities or non-physical artifacts. This shows that the method continues to produce physically realistic reconstructions, even when the data are compressed by more than $200\times$.

Taken together, these observations demonstrate that the discontinuous-DLS approach effectively preserves the essential flow dynamics and physical consistency as long as the reconstruction error remains within reasonable limits. For low to moderate target errors ($\leq 1\%$), the method produces reconstructions that closely match the original solution. At higher target errors (e.g., 5\%), some loss of small-scale detail is inevitable, yet the dominant flow features remain well captured. This trade-off between compression efficiency and reconstruction fidelity highlights the practical value of the method for both large- and small-scale fluid flow simulations, particularly in applications where storage savings are prioritized over exact preservation of fine gradients.

\section{Compression analysis on time-series data}

Building on the preceding assessment of spatial reconstruction quality and compression efficiency, this section examines the performance of the method on time-series data. In particular, the analysis focuses on how well key quantities such as kinetic energy (KE), turbulent kinetic energy (TKE), and power spectral density (PSD) are recovered from the compressed data.

\begin{figure}
    \centering
        \includegraphics[width=\linewidth]{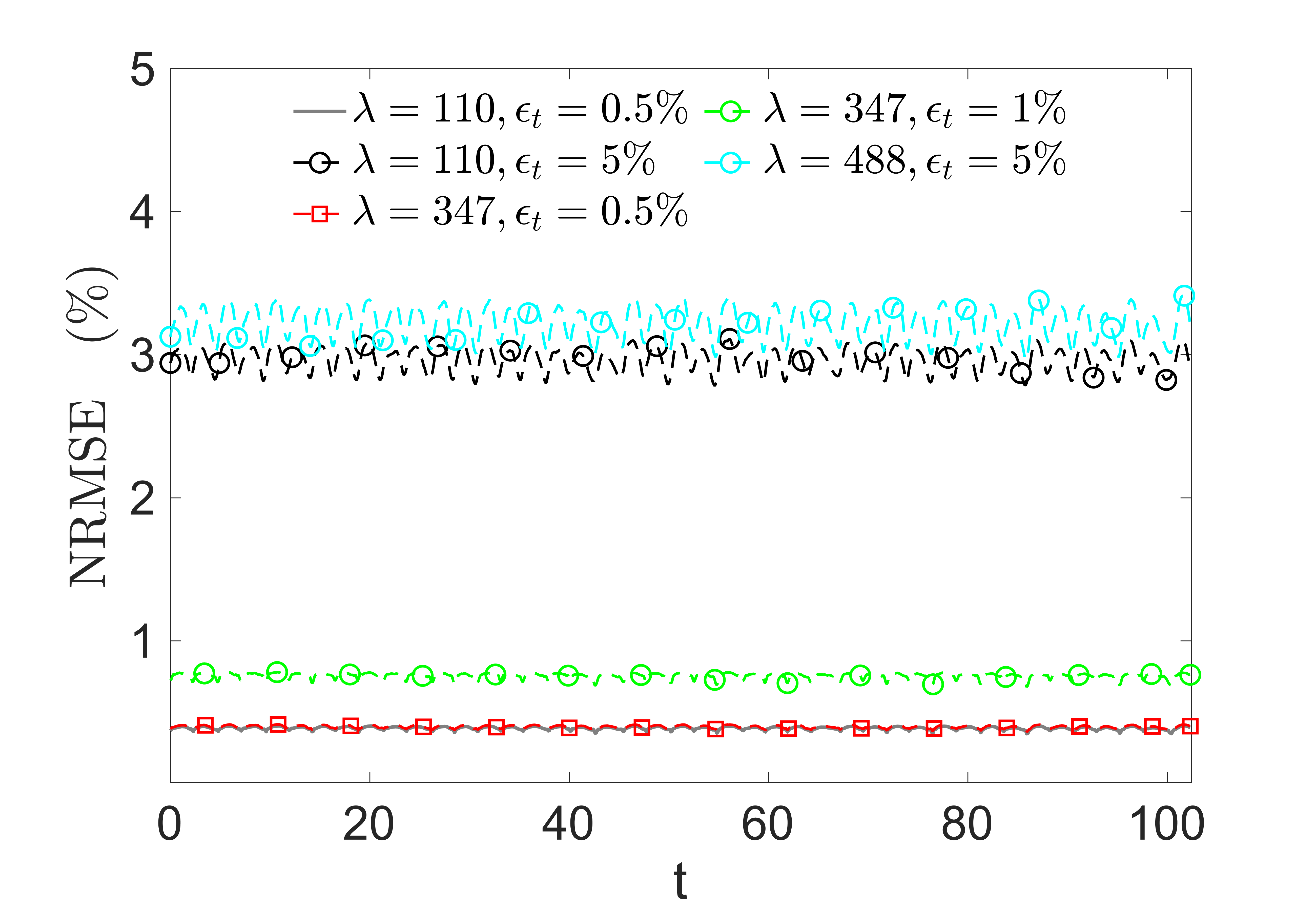}
    \caption{NRMSEs of the reconstructed solutions evaluated over 1024 snapshots spanning the first 102.4 time units of the flow past a cylinder dataset.}
    \label{fig:NRMSE_time}
\end{figure}

Figure \ref{fig:NRMSE_time} shows the temporal evolution of the normalized root mean square error (NRMSE) over the full reconstructed sequence of 1024 snapshots for five representative compression settings. The cases span three coarsening factors (110, 347, and 488) and multiple user-prescribed target error bounds. Here, the NRMSE serves as a global diagnostic to assess how closely the reconstruction adheres to the specified error tolerance over time.
For coarsening factors of 110 and 347 with a target error bound of 0.5\%, the observed NRMSE remains consistently below the prescribed limit throughout the entire time sequence, fluctuating between 0.35\% and 0.42\%. This indicates that the compression algorithm not only satisfies the user-defined error constraint, but does so uniformly in time, without evidence of error growth or temporal drift. The corresponding compression ratios of 46$\times$ and 67$\times$ further demonstrate that increasing the coarsening factor enables higher compression while remaining within the same error budget.

When the target error bound is relaxed to 1\% at a coarsening factor of 347, the NRMSE increases accordingly, taking values in the range of approximately 0.67\%–0.78\%. Importantly, the error remains well controlled and below the prescribed tolerance, reflecting the expected response of an error-bounded compression scheme. At the more aggressive target error of 5\%, the NRMSE rises to values around 3\%, accompanied by compression ratios approaching 200$\times$. A similar behavior is observed for the case with a coarsening factor of 488 and a target error of 5\%, where the NRMSE ranges from 2.96\% to 3.41\% with a compression ratio of 217$\times$. These results are consistent with the expected trade-off between compression efficiency and reconstruction accuracy. Notably, despite differences in absolute error levels, the NRMSE remains relatively steady over time in all cases, indicating that the error-bounded compression strategy preserves temporal consistency even at high compression ratios.

Overall, these results confirm that both the coarsening factor and the user-specified error tolerance play a central role in determining the compression outcome. Although the absolute reconstruction error varies across cases, the NRMSE remains temporally stable in all configurations, indicating that the prescribed error bound is consistently respected over the full time horizon. This temporal robustness is closely tied to the algorithm’s design, which leverages data-adaptive basis functions learned from the initial snapshot to capture the dominant flow structures and reuse them across subsequent time steps. As a result, the error-bounded framework remains reliable even under aggressive data reduction, making it well suited for long, time-dependent simulations where controlled error behavior is essential.

\begin{figure*}[h!]
    \centering
  \centering
    \begin{subfigure}[b]{0.45\linewidth}
        \centering
         \includegraphics[width=\linewidth]{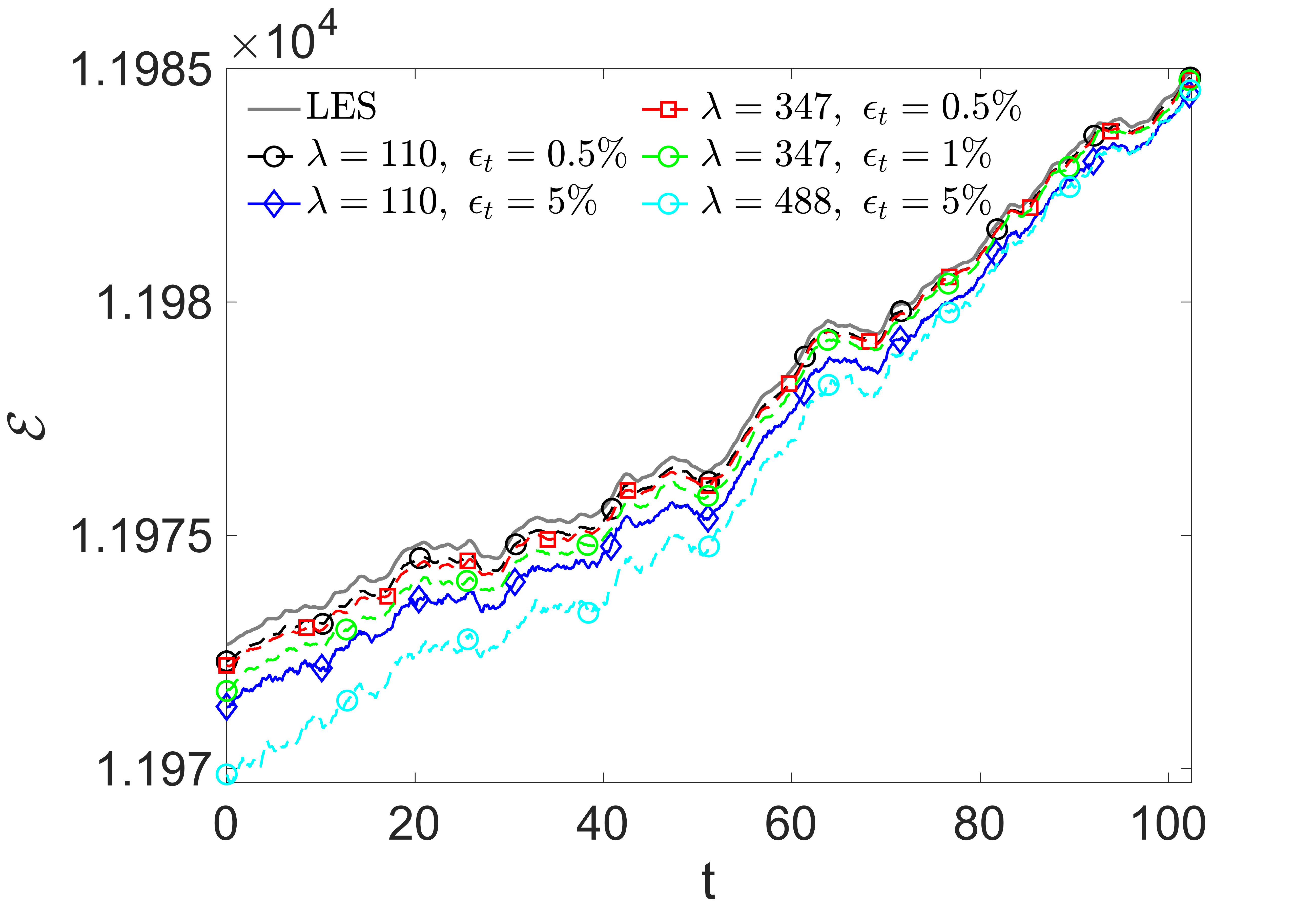}
        \caption*{(A) Kinetic energy}
    \end{subfigure}
    \begin{subfigure}[b]{0.45\linewidth}
        \centering
          \includegraphics[width=\linewidth]{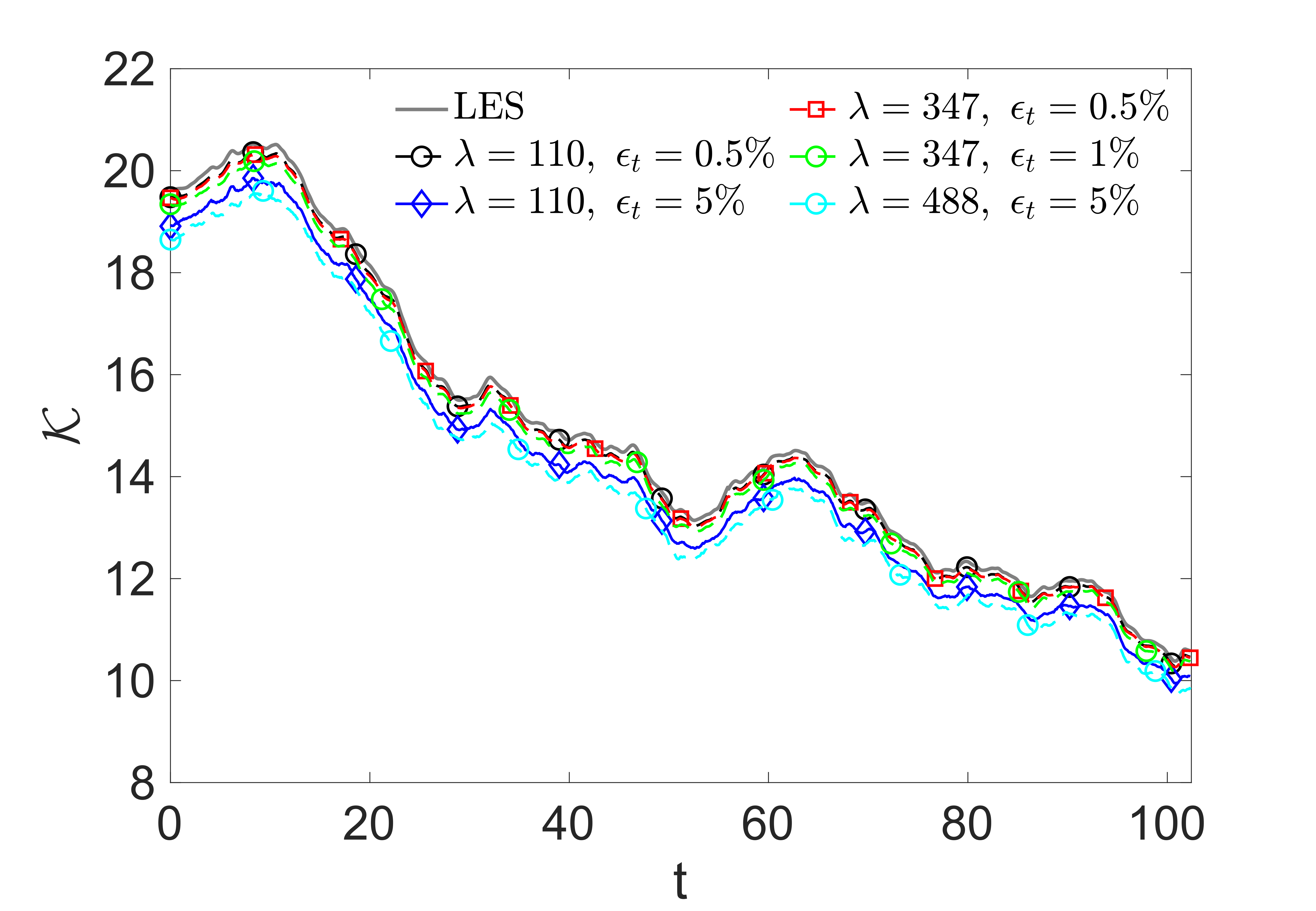}
        \caption*{(B) Turbulent kinetic energy}
    \end{subfigure}


    \caption{Comparison of instantaneous kinetic energy ($\mathcal{E}$) and turbulent kinetic energy ($\mathcal{K}$) from LES snapshots of flow past a cylinder over 102.4 time units, with the discontinuous-DLS approximations: (A) kinetic energy and  (B) turbulent kinetic energy.}
     \label{fig:KE_TKE}
\end{figure*}

Figure~\ref{fig:KE_TKE} shows the temporal evolution of the nondimensional kinetic energy ($\mathcal{E}$) and turbulent kinetic energy ($\mathcal{K}$) obtained from the reconstructed fields for the same five representative cases considered in Figure~\ref{fig:NRMSE_time}. These results are directly compared against the corresponding large eddy simulation (LES) data to assess the fidelity of the reconstructed flow fields.

Over the entire time interval, the reconstructed solutions exhibit strong agreement with the reference LES results, recovering more than 99.9\% of both $\mathcal{E}$ and $\mathcal{K}$. Notably, the method accurately reproduces not only the total kinetic energy associated with the mean flow but also the turbulent kinetic energy that reflects the contribution of unsteady, small-scale fluctuations. This level of agreement highlights the robustness of the compression approach in preserving the dominant energetic content of the flow, even when the data volume is substantially reduced.

\begin{figure*}[h!]
    \centering
  \centering
    \begin{subfigure}[b]{0.45\linewidth}
        \centering
         \includegraphics[width=\linewidth]{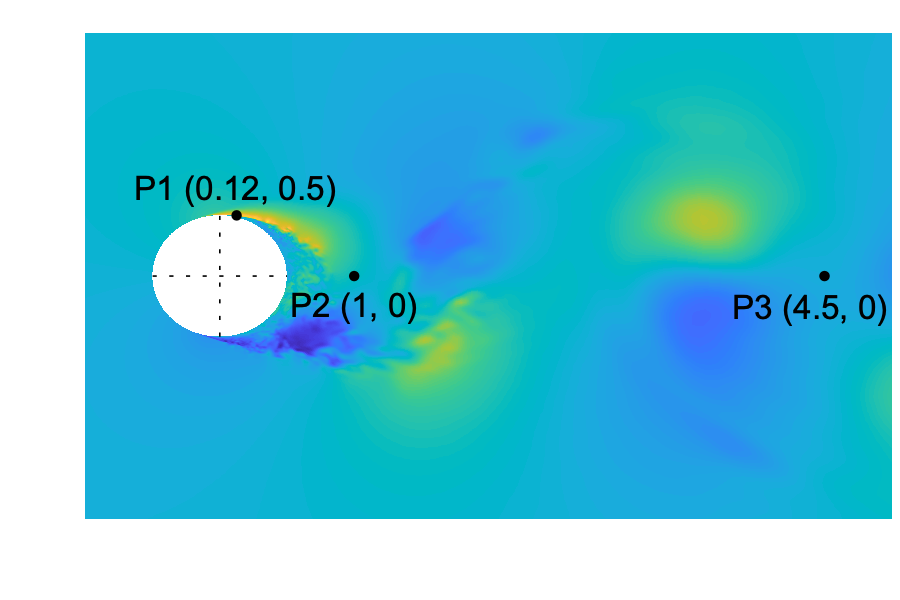}
        \caption*{(A) Probe locations}
    \end{subfigure}
    \begin{subfigure}[b]{0.45\linewidth}
        \centering
          \includegraphics[width=\linewidth]{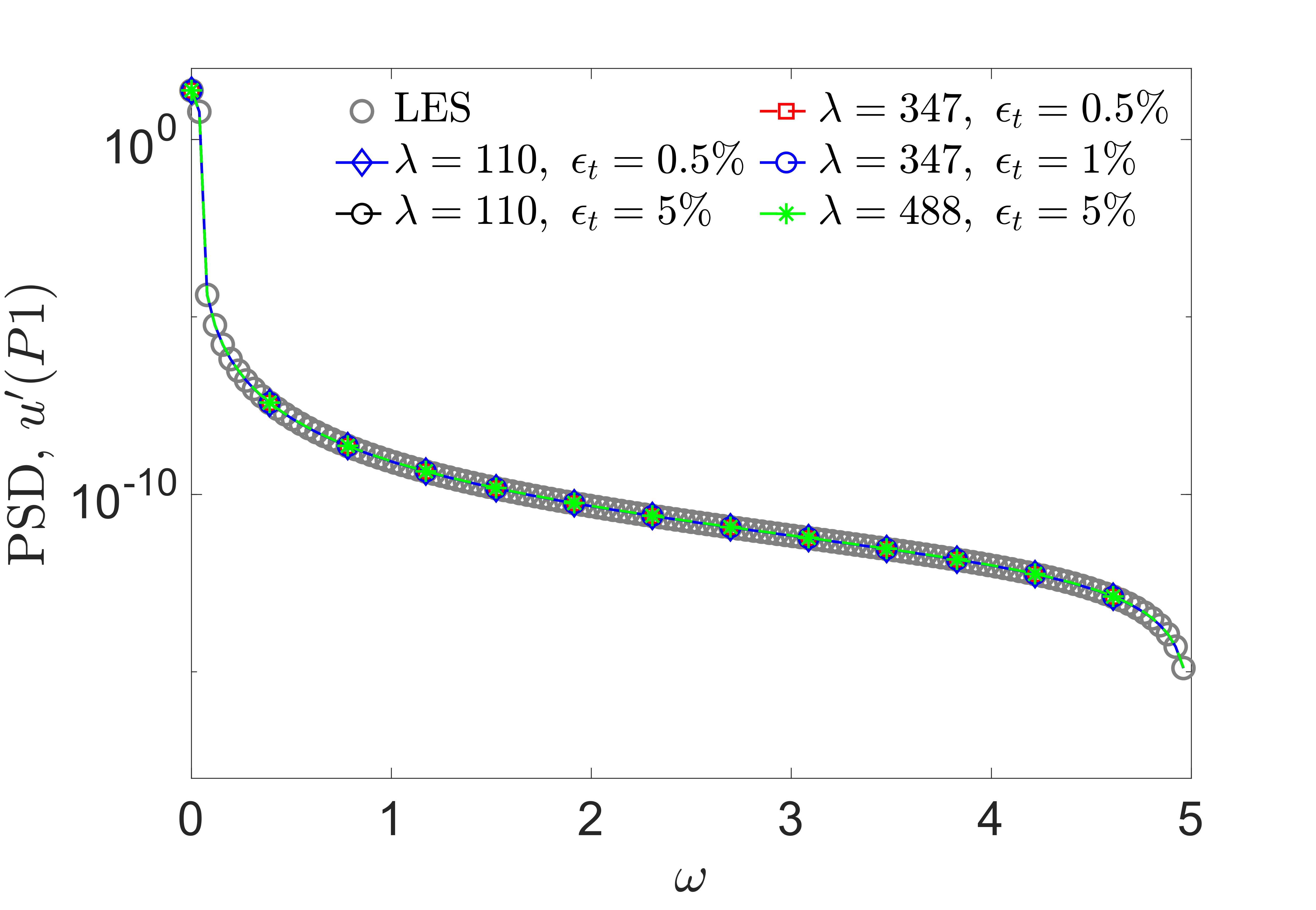}
        \caption*{(B) PSD of $u'$ at P1}
    \end{subfigure}


  \centering
    \begin{subfigure}[b]{0.45\linewidth}
        \centering
        \includegraphics[width=\linewidth]{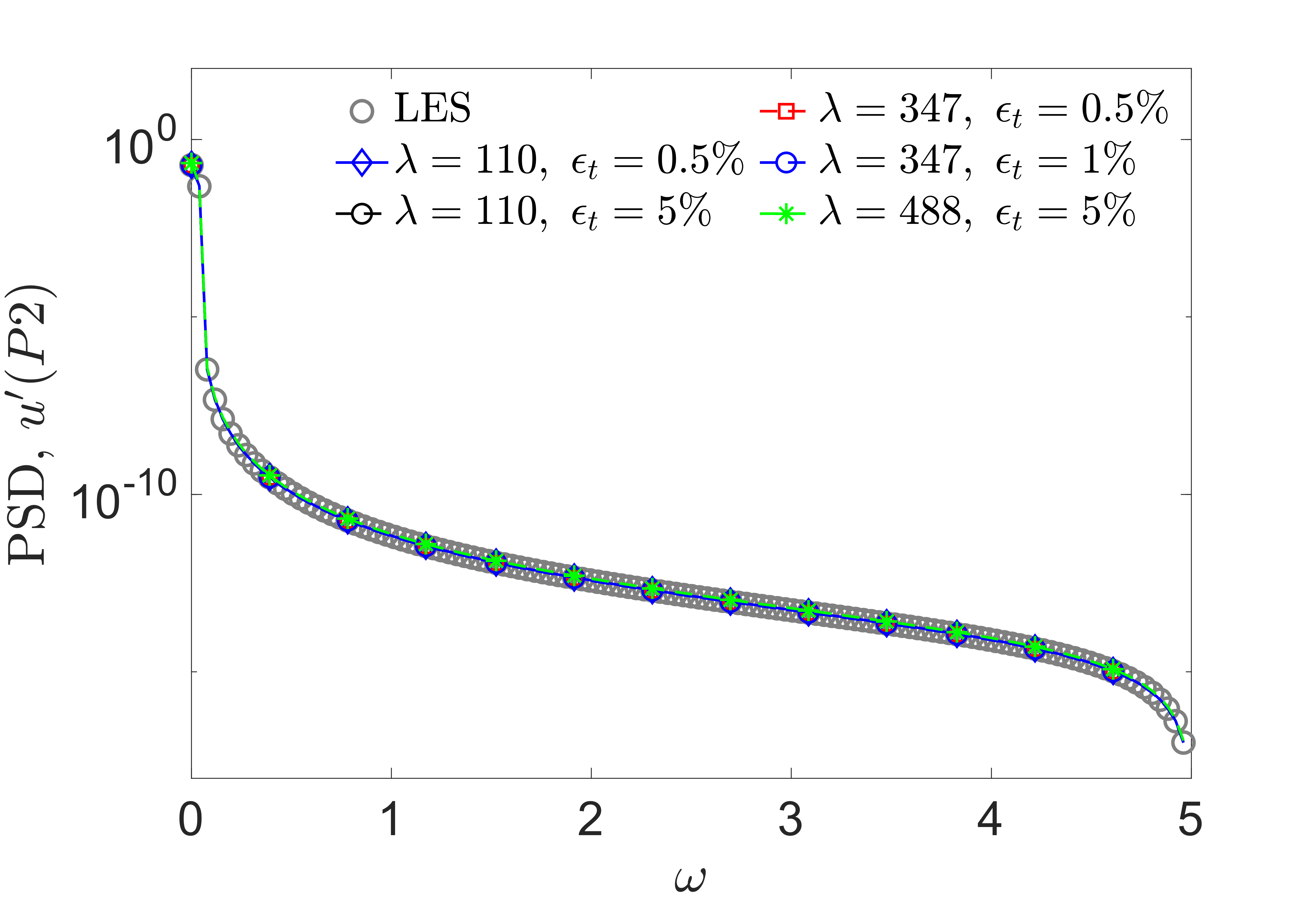}
        \caption*{(C) PSD of $u'$ at P2}
    \end{subfigure}
    \begin{subfigure}[b]{0.45\linewidth}
        \centering
         \includegraphics[width=\linewidth]{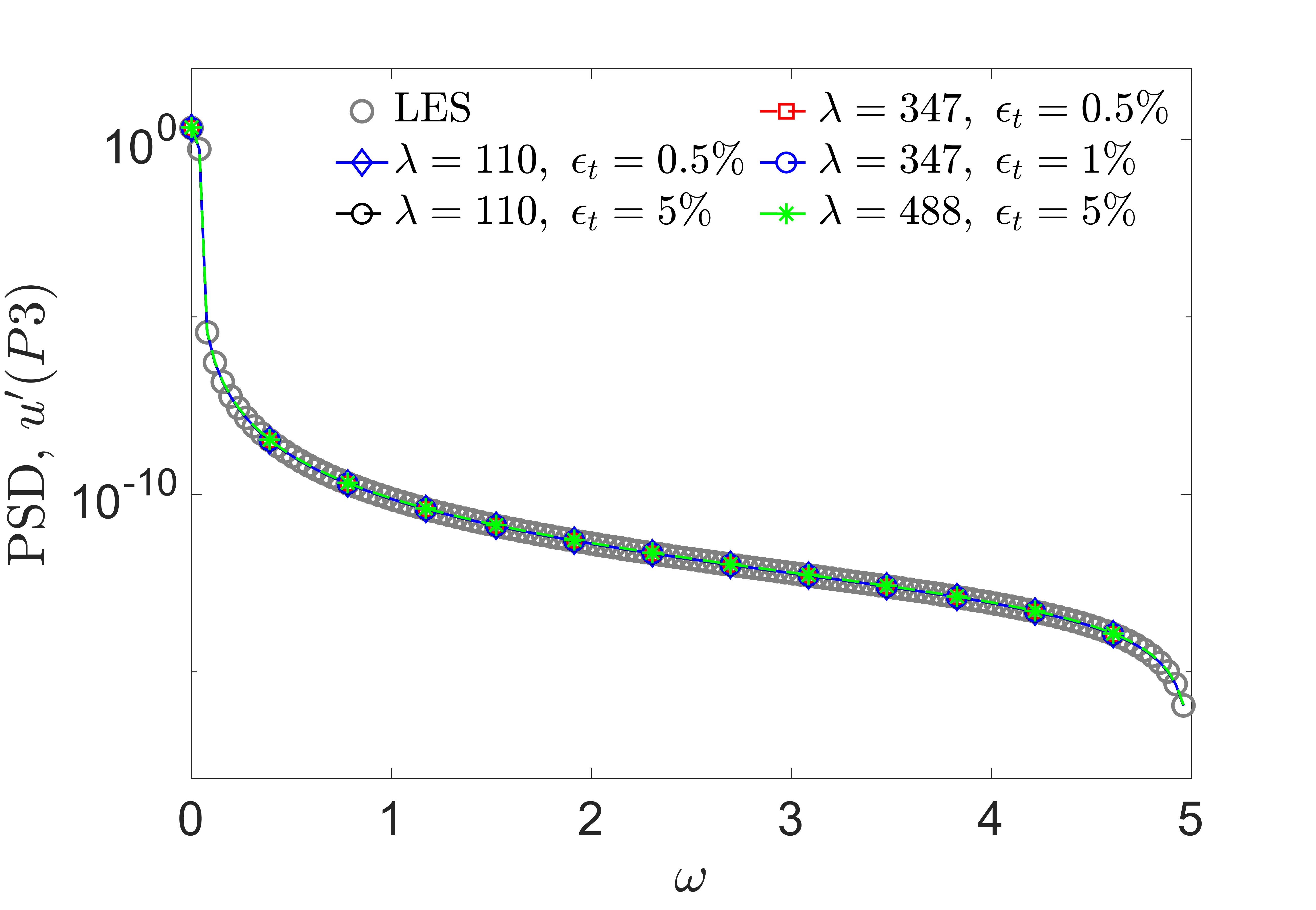}
        \caption*{(D) PSD of $u'$ at P3}
    \end{subfigure}


    \caption{Power spectral density (PSD) of the $u'$ velocity component over the first 102.4 time units of the flow past a cylinder, measured at probe locations P1–P3. (A) Probe locations, (B)
    PSD at P1, (C) PSD at P2, and (D) PSD at P3.}
      \label{fig:PSD}
\end{figure*}

To further assess the local and scale-dependent fidelity of the reconstructed solutions, power spectral densities (PSDs) of the velocity fluctuations are examined at three different probe locations. These include P1 $(0.12, 0.5)$ near the upper surface of the cylinder, P2 $(1, 0)$ in the near-wake region, and P3 $(4.5, 0)$ farther downstream in the wake. The selected probe points are intended to capture key flow features such as near-wall shear effects, vortex shedding dynamics, and far-wake turbulence, as illustrated in Figure~\ref{fig:PSD}(A). The PSDs are computed from the radial velocity fluctuation component, $u'$, extracted over the full time series of 1024 snapshots from both the reconstructed data and the reference LES data.

A comparison between the reconstructed and LES-based PSDs is shown in Figures~\ref{fig:PSD}(B–D) for the same five cases presented in Figures~\ref{fig:NRMSE_time} and \ref{fig:KE_TKE}. Across all three probe locations, the reconstructed spectra show excellent agreement with the LES results for all five cases, independent of the selected coarsening factor or target error level. Both the dominant frequencies and the overall spectral energy distributions are consistently recovered, demonstrating that the method preserves key temporal and frequency-domain characteristics of the flow. These results confirm the suitability of the proposed error-bounded compression approach for compressing large, unsteady datasets while retaining the essential physical and dynamical features of the underlying flow.



\subsection{Performance Analysis}

\begin{figure}[h!]
    \centering
         \includegraphics[width=\linewidth]{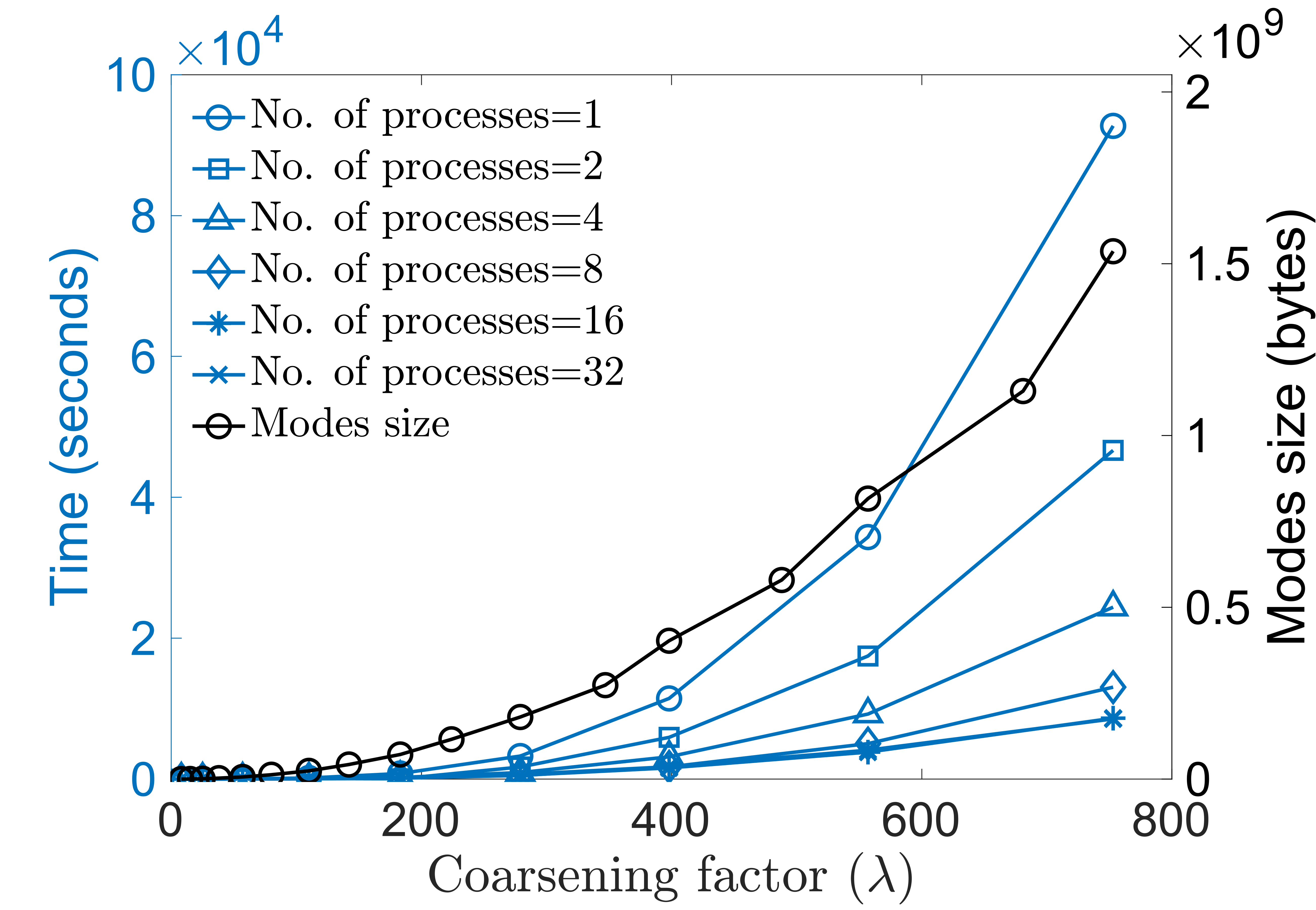}
    \caption{Learning time of basis functions and their storage cost as a function of coarsening factor $\lambda$ for compressing 1024 snapshots of 3D flow past a cylinder, with a total data size of $1.008 \times 10^{12} \, \text{bytes}$ (approximately 937.5 gigabytes).}
    \label{fig:learning_cost}
\end{figure}

This section evaluates the overall performance characteristics of the proposed compression approach, including the cost of basis construction, storage overhead, parallel scalability, and compression throughput. We first examine the compressor buildup stage, which corresponds to the one-time cost of learning the basis modes (i.e., the compression matrix) for the full dataset. The discussion focuses on how the time required to compute these modes varies with the coarsening level and processor count, as well as on the resulting size of the learned basis, which directly affects storage requirements and data transfer efficiency. In Figure~\ref{fig:learning_cost}, the left y-axis shows the time (in seconds) required to compute the modes, while the right y-axis shows the corresponding size of the basis modes in bytes.
The coarsening parameter $\lambda$ was varied over a range from 8 to 700. As anticipated, larger values of $\lambda$ significantly increase the computational cost due to the increase in the size of sample matrix and the associated SVD complexity. The buildup time exhibits superlinear growth with increasing $\lambda$. For example, on a single processor, the buildup time increases from approximately 5 seconds at $\lambda=8$ to over 92,000 seconds at $\lambda=700$.
Parallel performance was assessed using up to 32 processors. Substantial speedups were observed with increasing processor count, especially for moderate to high $\lambda$ values. However, the scalability was sublinear, and for lower $\lambda$ values, increasing the number of processors actually resulted in longer runtimes due to parallel overheads such as inter-process communication and load imbalance. For example, at $\lambda = 8$, the runtime increased from 1.50 seconds with 8 processors to 5.53 seconds with 32 processors, illustrating the inefficiency of fine-grained parallelism for small-scale problems. The reported timings also include the cost of reading sampling data and writing the output basis modes. It is worth noting that for a fixed coarsening level, the learning cost of the basis is constant and relatively low compared to the total compression time. A detailed analysis of compression throughput is also presented in the next figure.

We also evaluated the storage cost associated with the computed basis modes. As $\lambda$ increases, the size of the basis modes grows, starting from approximately 100 MB at $\lambda = 8$ to about 1.5 GB at $\lambda = 700$, based on a total dataset size of \(1.008 \times 10^{12}\) bytes (approximately 937.5 GB). Importantly, both the learning cost and storage requirements remain unchanged when varying the global target error, since the basis is learned from a coarsened grid using SVD and captures the dominant global structures of the dataset. Only the compression and reconstruction phases are influenced by the target error.  In practice, selecting an appropriate $\lambda$ requires balancing compression quality, computation time, and available storage resources.

\begin{figure*}[h!]
    \centering
  \centering
    \begin{subfigure}[b]{0.45\linewidth}
        \centering
         \includegraphics[width=\linewidth]{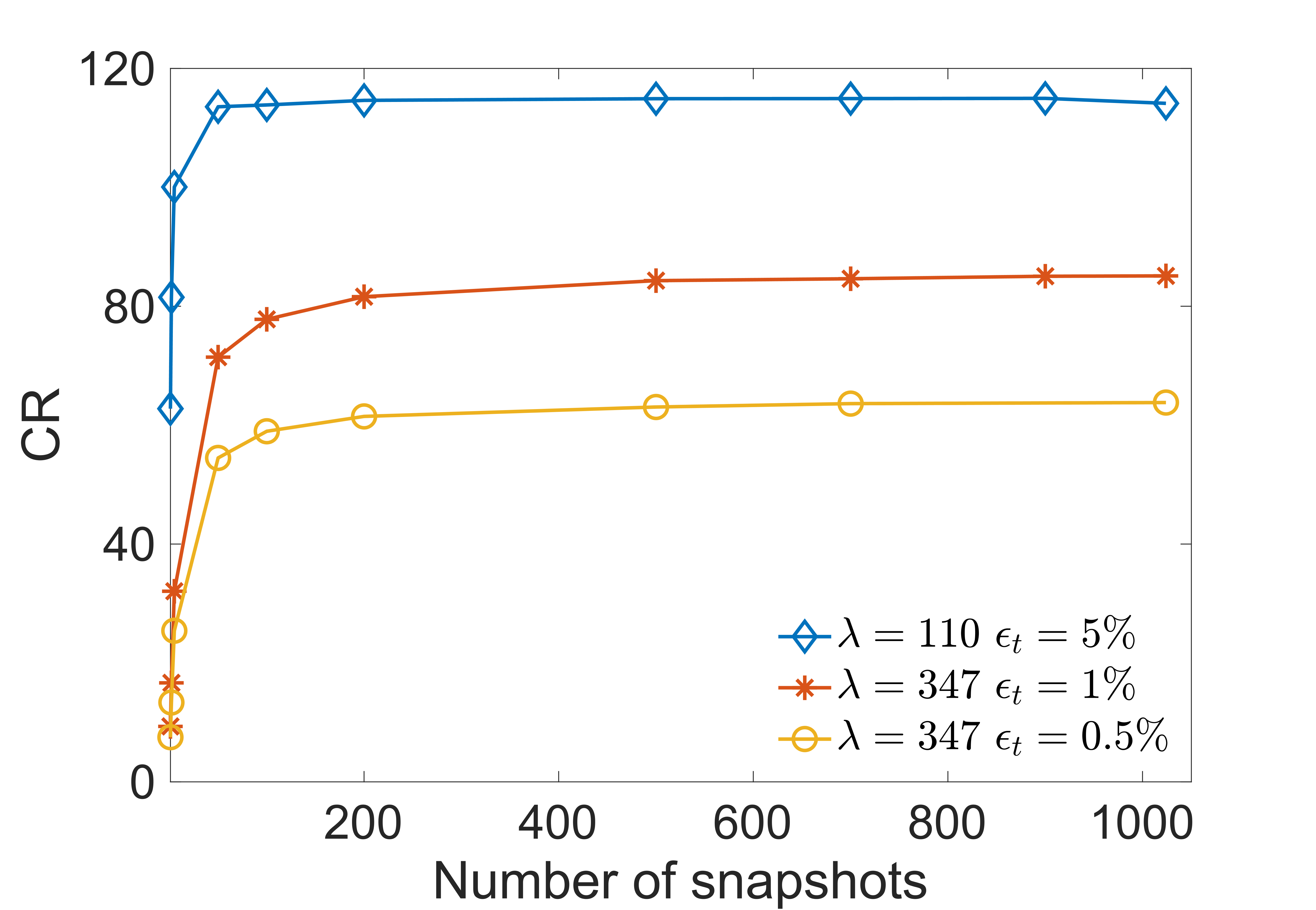}
        \caption*{(A)}
    \end{subfigure}
    \begin{subfigure}[b]{0.45\linewidth}
        \centering
          \includegraphics[width=\linewidth]{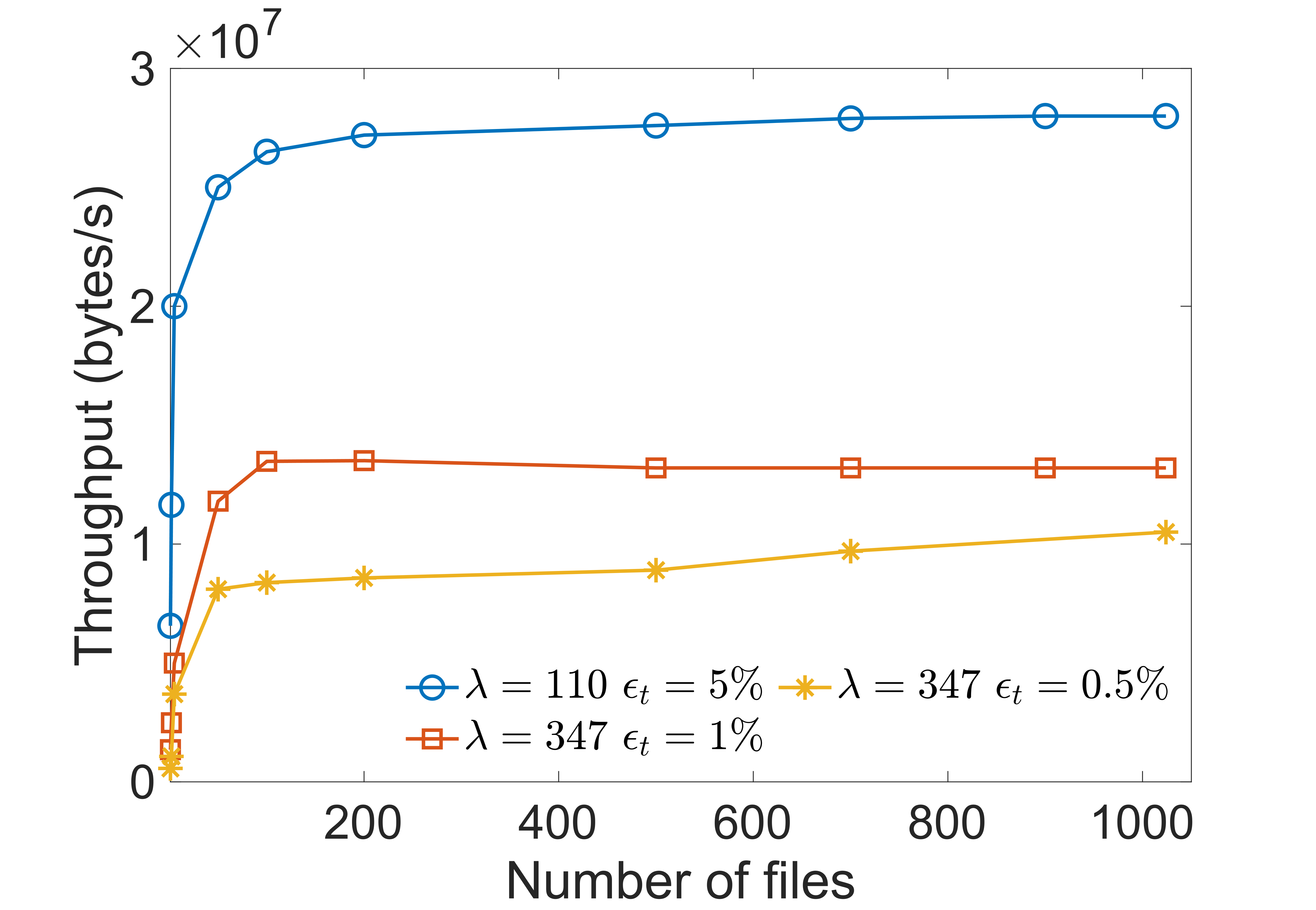}
        \caption*{(B)}
    \end{subfigure}


    \caption{Compression and throughput analysis for varying snapshot counts:
           (A) compression ratio vs. number of snapshots: the relationship between the compression ratio and the number of snapshots across different compression cases, and (B) compression throughput vs. number of snapshots: the achieved compression throughput as a function of the number of snapshots,  using parallel processing on a single node with 48 cores and 250GB memory.}
           \label{fig:throughput}
\end{figure*}



We next examine the compression throughput of the proposed algorithm by varying the number of snapshots and measuring the corresponding compression ratios and time taken. Throughput was computed as the ratio of the total input size to the compression time, offering a measure of how efficiently data can be compressed over time. Three different parametric settings were considered: (i) $\lambda = 110$, $\epsilon_t = 5\%$, (ii) $\lambda = 347$, $\epsilon_t = 1\%$, and (iii) $\lambda = 347$, $\epsilon_t = 0.5\%$.
%
As shown in Figure~\ref{fig:throughput} (a), the compression ratio increases with the number of snapshots for all configurations, eventually saturating as the number of samples becomes sufficient to capture the dominant features of the dataset. The configuration using the largest target error, $\epsilon_t = 5\%$, yields the highest compression ratio, exceeding $110\times$ when more than 500 snapshots are compressed. In contrast, stricter error tolerances lead to lower compression ratios, as a larger number of coefficients must be retained to satisfy the prescribed accuracy requirements. The corresponding compression throughput is shown in Figure~\ref{fig:throughput}(b). For the case with $\lambda = 110$ and $\epsilon_t = 5\%$, the throughput reaches a maximum of approximately 27 MB/s for datasets containing 500 to 1024 snapshots, indicating improved efficiency as the dataset size increases. Throughput was generally lower for tighter error thresholds and larger coarsening factors due to increased computational cost. For example, at $\lambda = 347$, $\epsilon_t = 1\%$, the throughput reached around 13 MB/s, while further reducing $\epsilon_t$ to 0.5\% resulted in a peak throughput below 10 MB/s. These results suggest that selecting a coarsening factor greater than 120 provides a favorable compromise between compression efficiency and computational cost.

Overall, the observed trends highlight the inherent trade-off between compression speed and reconstruction accuracy. Enforcing stricter accuracy requirements (smaller $\epsilon_t$) increases the computational cost and reduces throughput, whereas allowing larger reconstruction errors accelerates compression at the expense of fidelity. Consequently, the choice of coarsening factor $\lambda$ and target error threshold $\epsilon_t$ should be guided by application-specific priorities, depending on whether computational efficiency or reconstruction quality is the primary concern.

\section{Conclusion and Future Perspectives}

This work presented a discontinuous extension of the Data-Informed Local Subspaces (DLS) framework for error-bounded lossy compression of large-scale scientific datasets. By relaxing the global $C^0$
continuity constraint, the proposed approach enables localized control of reconstruction error while retaining the data-adaptive characteristics of DLS. Although the numerical experiments in this study focused on three-dimensional flow past a cylinder, the underlying formulation is independent of the specific physics and is readily applicable to structured-grid data arising in a broad range of scientific applications.

The numerical results indicate that the discontinuous DLS approach can achieve substantial compression while maintaining user-specified error bounds. Across a range of coarsening factors and user-defined error levels, the method produces reconstructed fields that remain physically consistent with the original data. Compared with widely used error-bounded compressors such as SZ and MGARD, discontinuous DLS shows competitive performance, particularly at lower error tolerances where preserving localized structure becomes more challenging for data-agnostic compression methods. 

To evaluate reconstruction quality beyond standard error metrics, several physically meaningful quantities were examined. Kinetic energy, turbulent kinetic energy, vorticity, and power spectral density were all analyzed using the reconstructed fields. These diagnostics show that both global energy content and small-scale dynamical features are well preserved after compression. In addition, the time-dependent analysis demonstrates that the method maintains temporal consistency over long sequences of snapshots and shows no evidence of error accumulation. The performance analysis illustrates the practical behavior of the proposed approach. Although the cost of learning the data-driven basis increases with the coarsening level, this one-time cost can be amortized over long time-series datasets. Parallel and throughput results indicate that the localized formulation enables effective scaling for moderate to large problem sizes, while revealing the expected trade-off between compression efficiency, computational cost, and reconstruction accuracy.

As expected, the cost of constructing the data-informed basis increases with coarsening level; however, this expense is incurred only once and can be amortized efficiently over large time-series datasets. Parallel experiments indicate that the patch-based structure of discontinuous DLS is well suited to distributed-memory execution for moderate to large problem sizes. Throughput results also highlight a predictable trade-off between computational cost and compression efficiency as the error tolerance is tightened, providing users with clear control over accuracy–performance balance. Future work will focus on improving the flexibility and scalability of the proposed framework. Planned efforts include support for multiple error bounds, improved efficiency at low error tolerances, and extensions to unstructured and adaptive meshes. Another important direction is the integration of discontinuous DLS into in-situ compression workflows for large-scale simulations, where data movement and storage costs are critical.

\section*{Acknowledgments}
The authors gratefully acknowledge support from the NASA SBIR Phase II grant 80NSSC23PB560, with Dr. Sharad Gaveli as Technical Monitor, and the DOE STTR Phase II grant DE-SC0022806, with David Henderson. All the computations have been carried out on the University of Central Florida’s ARCC Stokes cluster.

\vfill

%

\end{document}